# Non-magnetic spin splitting driven by spin-valley-layer coupling in multilayer $WSe_2$

Min-Gue Kim[1,2], Min-Sik Kim[1,2], Kenji Watanabe[3], Takashi Taniguchi[4], Ju-Jin Kim[1*], and Myung-Ho Bae[2,5*]

[1]Department of Physics, Jeonbuk National University, Jeonju 54896, Republic of Korea

[2]Korea Research Institute of Standards and Science, Daejeon 34113, Republic of Korea

[3] Research Center for Electronic and Optical Materials, National Institute for Materials Science, 1-1 Namiki, Tsukuba 305-0044, Japan

[4] Research Center for Materials Nanoarchitectonics, National Institute for Materials Science, 1-1 Namiki, Tsukuba 305-0044, Japan

[5]KAIST Graduate School of Quantum Science and Technology, Korea Advanced Institute of Science and Technology, Daejeon 34141, Republic of Korea

*e-mail: jujinkim@chonbuk.ac.kr, mhbae@kriss.re.kr

## Abstract

Transition metal dichalcogenides provide a platform for exploring spin-valley physics, offering a promising approach to electric-field-driven spin control for low-power spintronic and quantum devices. Here, we demonstrate electric-field-induced spin splitting in the Q and Q' valleys of multilayer n-type $WSe_2$ using quantum-point-contact spectroscopy. Systematic modulations in four distinct conductance quantization steps, providing direct evidence of spin-valley-layer coupling-driven spin-resolved density of states, are achieved by tuning the out-of-plane gate voltage. Notably, the electric-field-induced spin splitting significantly dominate the magnetic-field-induced valley-Zeeman effect (i.e., ~7 meV for a displacement field change of

~0.08 V/nm vs. ~2 meV for a magnetic field of $B$ = 9 T), demonstrating a powerful, non-magnetic manipulation of spin states. This ability to manipulate spin states by gate voltage is crucial for advancing next-generation low-power spintronic and quantum information technologies.

## Main text

Transition metal dichalcogenides (TMDCs), such as $MoS_2$ and $WSe_2$, have emerged as promising platforms for valleytronics due to their unique electronic band structures. In monolayer TMDCs, valleys at the K and K' points of the Brillouin zone serve as distinct degrees of freedom for information encoding[1,2,3,4,5,6,7]. These valleys are inherently coupled to spin via strong spin-orbit coupling, leading to spin-valley locking, where the K and K' valleys exhibit opposite spin polarizations while maintaining degeneracy protected by time-reversal symmetry[8]. In multilayer TMDCs, similar spin-valley locking behavior has been observed within the Q and Q' valleys[9]. Recent studies have demonstrated gate-voltage-induced Zeeman-type spin splitting in multilayer TMDCs, where an external electric field polarizes spins layer-by-layer without breaking time-reversal symmetry, thus without selective valley manipulation[10]. Importantly, multilayer systems provide an additional layer-related degree of freedom through spin-valley-layer (SVL) coupling[1,11,12]. In such systems, the electronic states in each valley exhibit valley-contrasted layer polarization, enabling direct coupling between valley degrees of freedom and an applied gate electric field.

Meanwhile, recent investigations using quantum-point-contact (QPC) spectroscopy have demonstrated spin-resolved quantized conductance plateaus in n-type $MoS_2$[13,14,15] and p-type $WSe_2$[16,17,18], regardless of the number of layers, suggesting the formation of valley- and spin-polarized states within the QPC constriction. The observation of spin splitting in QPC spectral data further implies the influence of SVL coupling within the confined region. However, a definitive demonstration of electrically induced, non-magnetic spin splitting and its systematic control via the SVL coupling remain elusive. Harnessing the SVL degree of freedom within a confined QPC channel offers a unique route toward achieving gate-tunable spin control without an external magnetic field.

In this study, we report unambiguous evidence of the systematic control of spin splitting via SVL coupling through QPC transport measurements in n-type multilayer $WSe_2$. We show that the gate voltage systematically tunes four distinct spin-resolved subbands, in quantitative agreement with an electric-field-induced SVL coupling model. Our findings demonstrate electric-field-based control of spin states, achieving a splitting of ~7 meV for a displacement field change of only ~0.08 V/nm. This result is notably superior to the magnetic-field-induced Zeeman splitting, which reaches only ~2 meV at 9 T. By establishing QPCs as a robust platform for engineering spin-valley degrees of freedom, this work represents a significant step toward the realization of practical, low-power spintronic and valleytronic devices.

**Electrical characterization of devices**

The left panel of Fig. 1a shows a schematic of the $WSe_2$ QPC device architecture employed in this study. This structure comprises a global Si back-gate electrode with a 500 nm-thick $SiO_2$ layer, a pair of split gates for QPC definition (see the right panel of Fig. 1a), a hexagonal boron nitride (*h*-BN) dielectric layer, a multilayer $WSe_2$ channel, and In contacts[19]. We characterized two multilayer $WSe_2$ devices, Device #1 ($N_L$ = 14, $L$ = 11.5 μm, and $W$ = 5 μm) and Device #2 ($N_L$ = 5, $L$ = 4 μm, and $W$ = 3 μm), as shown in the optical images in the left and right panels of Fig. 1b, respectively. Here, $L$ and $W$ represent the channel length and width, respectively, and $N_L$ denotes the layer number of $WSe_2$, as confirmed by atomic force microscopy (detailed fabrication processes and measurement configurations are provided in Supplementary Note 1). Fig. 1c, d present the current ($I$) as a function of back-gate voltage ($V_{BG}$) for Device #1 and Device #2, respectively, across a range of temperatures ($T$s). Both devices exhibit characteristic n-type transport behavior, which is expected for In contacts on multilayer $WSe_2$ due to the close alignment between the In work function (~4.1 eV) and the $WSe_2$ electron affinity (~4 eV)[20]. For both devices, the slope of the $I$-$V_{BG}$ curve becomes steeper as $T$ decreases, indicating an enhancement in field-effect mobility ($\mu$) at lower $T$s. In Figure 1e, the $T$ dependence of mobility ($\mu$) for Device #1 (green circles) and Device #2 (black squares) is presented. The mobility was calculated using the relation $\mu = \frac{L}{W}\frac{1}{C_{BG}V_{\mathrm{SD}}}\frac{\mathrm{d}I}{\mathrm{d}V_{BG}}$ with a source-drain voltage ($V_{SD}$) of 20 mV for both devices. Here, $C_{BG} = \left(\frac{1}{C_{\mathrm{SiO_2}}} + \frac{1}{C_{h-\mathrm{BN}}}\right)^{-1}$, where $C_{\mathrm{SiO_2}}$ and $C_{h-\mathrm{BN}}$ represent the capacitance of the $SiO_2$ and $h$-BN substrate layers, respectively. The $\mu$ values were extracted at $V_{BG}$ = 45 V for Device #1 and 70 V for Device #2, where the mobility remains relatively independent of $V_{BG}$. For both devices, $\mu$ increases as $T$ decrease and saturates at $T$ < 30 K, reaching 680 and 185 $cm^2V^{-1}s^{-1}$ for Device #1 and Device #2, respectively, at $T$ = 4 K. The solid curves in Fig. 1e represent fits to the Matthiessen's rule $\mu_{\mathrm{tot}}(T) = \left[\frac{1}{\mu_{\mathrm{imp}}} + \frac{1}{\mu_{\mathrm{ph}}(T)}\right]^{-1}$, where the total mobility ($\mu_{\mathrm{tot}}$) incorporates both $T$-independent impurity scattering ($\mu_{\mathrm{imp}}$) and $T$-dependent phonon scattering ($\mu_{\mathrm{ph}}$). At the lowest base $T$ (~ 4 K), where acoustic and optical phonon scatterings are effectively frozen out, $\mu_{\mathrm{tot}}$ saturates to the impurity-limited mobility, yielding $\mu_{\mathrm{imp}} \approx$ 680 and 185 $cm^2V^{-1}s^{-1}$ for Device #1 and Device #2, respectively. Phonon scattering is described by $\mu_{\mathrm{ph}} \propto T^{-\alpha}$, with extracted exponents of $\alpha$ = 2.2 for Device #1 and 1.8 for Device #2 for the fitting, as indicated by the dashed lines in Fig. 1e. The $T$ scaling of mobility in 2D TMDCs is not a universal constant; rather, it is highly sensitive to the intrinsic lattice dynamics, channel thickness, and the dielectric screening environment. Our extracted values of $\alpha$ values reflect the distinct phonon dispersion relations and heavier atomic masses of the $WSe_2$ lattice, which inherently differ from those of $MoS_2$[21]. Furthermore, the asymmetric dielectric environment, where the $WSe_2$ channel is supported by a bottom *h*-BN layer, modifies the screening of remote optical phonons compared to fully encapsulated structures[21]. Thus, this exponent represents the intrinsic electron-phonon coupling characteristics unique to our *h*-BN-supported multilayer $WSe_2$ configuration.

**Origin of diamond structures in transconductance map**

Figure 2a shows the zero-bias differential conductance ($G$) of Device #1 as a function of the split-gate voltage ($V_{QPC3}$) at $V_{BG}$ values ranging from 68 V (the leftmost curve) to 50 V in increments of 0.5 V and $T = 20$ K. The conductance is plotted in units of $2e^2/h$, where $e$ and $h$ are the elementary charge and Plank's constant, respectively. Several distinct groups of conductance steps are observed, which can be traced across the varying $V_{BG}$ values, such as the orange, green, and grey groups. To investigate the physical origin of these steps, we examined the transconductance map ($dG/dV_{QPC3}$) as a function of $V_{SD}$ and $V_{QPC3}$ at $V_{BG} = 60$ V (Fig. 2b), which corresponds to the red curve in Fig. 2a. The transconductance map reveals a well-defined checkerboard pattern, highlighted by white guide lines (also see Supplementary Note 2). Figure 2c illustrates $G$ as a function of $V_{SD}$ for various $V_{QPC3}$ values ranging from −1.9 V (top curve) to −2.7 V in 10 mV increments. Conductance plateau groups emerge at $V_{SD} = 0$ mV (orange and green arrows), ~11 mV (blue arrow), and ~−11 mV (brown arrow), with corresponding regions identified by color-coded points in Fig. 2b. While the grey conductance group is not clearly distinct in Fig. 2c, it is well-resolved in the transconductance map (Fig. 2b, grey arrow).

$T$-dependent measurements were performed on Device #1 at $V_{BG}$ = 60 V (Fig. 2d). The relatively smooth conductance plateaus observed at $T \geq$ 20 K in the green and grey shaded regions evolve into pronounced peak-dip conductance fluctuations at $T \leq$ 10 K. Despite these fluctuations, the macroscopic baseline conductance remains largely $T$-independent, characteristic of metallic behavior. In contrast, the conductance steps in the orange-shaded region decrease with falling $T$, exhibiting insulting-like behavior. The green- and orange-shaded conductance plateaus observed at $T$ = 20 K in Fig. 2d correspond to green and orange dashed diamonds in Fig. 2b, respectively, which exhibit energy scales of ~7.5 meV, as indicated by white arrows. If these features originated from a Coulomb blockade effect, this energy scale would correspond to the charging energy, and the local conductance peaks observed at $T \leq$ 10 K would represent resonant tunneling through a quantum dot. However, a charging energy corresponding to ~7.5 meV significantly exceeds the thermal energy at 4 K (~0.3 meV), thus, it would inevitably induce a robust Coulomb blockade, characterized by near-zero conductance, between the resonant peaks. This expectation sharply contradicts our observation of merely a shallow conductance dip between the two peaks. Such a discrepancy indicates that the quantum-dot scenario is inadequate for explaining the observed checkerboard pattern. Instead, the spectrum strongly reflects 1D subband formation, allowing us to directly evaluate the subband energy and the effective spatial length scales of the QPC electrostatic confinement.

For Device #1, the lithographic width and length of the QPC are ~140 nm and ~200 nm, respectively (see the right panel of Fig. 1a). However, the actual electronic path is defined not by these physical gate boundaries, but by the electrostatic confinement potential induced by the split gates. When a strong negative voltage is applied to the split gates, i.e., near pinch-off condition, fringing electric fields penetrate through the $h$-BN dielectric, creating extensive depletion regions within the $WSe_2$ layer. These depletion regions extend laterally far beyond the physical edges of the metal gates. This intense "electrostatic squeezing" effect drastically narrows the conducting path. According to the Büttiker saddle-point potential model[22], the effective width ($W_{\mathrm{eff}}$) of QPC should be comparable to half of the Fermi wavelength ($\lambda_F/2$), corresponding to the ground level of the subbands formed within a harmonic potential. Supplementary Note 3 details the experimentally obtained physical parameters and length scales of the examined devices.

For Device #1, we extracted a Fermi wavelength $\lambda_F \sim 17.4$ nm, a mean free path $l_e \sim 40$ nm, the effective width $W_{\mathrm{eff}} \sim 5.6$ nm, and an effective length $L_{\mathrm{eff}} = 11.2$ - $28$ nm. The observation of a distinct subband energy gap is primarily attributed to the fact that $W_{\mathrm{eff}}$ is comparable to $\lambda_F/2$. This condition facilitates the clean isolation of the fundamental 1D QPC confinement subband. Although the condition $L_{\mathrm{eff}} \leq l_e$ typically implies ballistic transport, the specific transport signatures observed at $T \leq 10$ K necessitate a more rigorous classification (see Fig. 2d). In this low-$T$ regime, pronounced peak-dip conductance fluctuations appear superimposed on the quantized plateaus. This phenomenon occurs as the phase coherence length ($l_\phi$) extends significantly, far exceeding the channel length ($l_\phi \gg L_{\mathrm{eff}}$) (see Supplementary Note 3). Consequently, even minimal residual impurities or local potential fluctuations in the vicinity of the channel can act as phase-coherent scattering centers. The observed peak-dip features are clear manifestations of quantum resonant backscattering, where electrons undergo multiple coherent reflections. These reflections lead to quantum interference that slightly degrades the perfect forward transmission ($Tr < 1$). Therefore, by accounting for both the length-scale hierarchy and the low-$T$ interference effects, we classify the transport mechanism of Device #1 as operating in the phase-coherent quasi-ballistic regime.

Based on preceding analysis, we attribute the orange and green shaded regions in Fig. 2d to the $e^2/h$ and $2e^2/h$ conductance plateaus, respectively. The significant suppression of conductance to approximately ~$0.2e^2/h$ for the initial plateau, relative to the ideal $e^2/h$, is likely due to residual disorder potentials and strong backscattering. In the orange shaded region, the conductance step exhibits an insulating trend, with values increasing as a function of $T$. Arrhenius analysis for this specific regime (upper panel, Fig. 2e) yields an activation energy of $E_a$ = 2.2 meV. This activation barrier implies that as the Fermi level approaches the channel depletion point, electrons encounter the background potential roughness of the channel.

While the absolute magnitude of the conductance steps is attenuated by disorder-limited transmission, we emphasize that the fundamental 1D macroscopic quantization remains remarkably preserved. The definitive evidence for QPC subband formation lies not in the absolute plateau height, but in the robust finite-bias diamond structures, representing subband energy spacings, revealed in the transconductance maps. These structures remain fully intact regardless of longitudinal backscattering. Indeed, the conductance steps observed in the orange and green regions of Fig. 2d directly correspond to the orange and green dashed diamond structures in Fig. 2b, as aforementioned.

**Spin splitting of subbands in Device #1**

Figure 3a presents transconductance maps ($\mathrm{d}G/\mathrm{d}V_{\mathrm{QPC3}}$) for Device #1 as a function of $V_{\mathrm{SD}}$ and $V_{\mathrm{QPC3}}$ for various $V_{\mathrm{BG}}$ values ranging from 44 V to 60 V at $T$ = 4 K and $V_{\mathrm{QPC2}}$ = 4 V. White lines are provided as guides for the eye to highlight the subband structure, following established spectroscopic rules (see Supplementary Note 2). In the map for $V_{\mathrm{BG}}$ = 60 V (Fig. 3a, rightmost panel), the $\Delta_1$ and $\Delta_{1/2}$ energy scales correspond to the green and orange diamonds identified at $T$ = 20 K in Fig. 2b, respectively, where the index $n$ in $\Delta_n$ denotes the corresponding subband with $G = nG_0$. Notably, the energy scales extracted along the $V_{QPC3}$ axis ($\delta V_{QPC3}$), indicated by the bidirectional arrows, exhibit contrasting dependencies on $V_{BG}$. As shown in Fig. 3b, $\Delta_1$ increases while $\Delta_{1/2}$ decreases as $V_{BG}$ is raised. However, when evaluating the energy scales from the horizontal $V_{SD}$ axis ($\delta V_{SD}^{+}$, indicated by the bidirectional arrows in Fig. 3a, second panel), both $\Delta_{1/2}$ and $\Delta_1$ decrease with increasing $V_{BG}$, albeit with differing slopes (Fig. 3c) (also see Supplementary Note 4 for $\delta V_{SD}^{-}$). While the behavior of $\Delta_{1/2}$ is consistent across both the $\delta V_{QPC3}$ and $\delta V_{SD}^{+}$ scales, $\Delta_1$ exhibits an inconsistent scaling between the two axes.

We attribute this inconsistent scaling of $\Delta_1$ in Device #1 to its relatively large thickness ($N_L$ = 14). In such thick multilayer structures, the two-dimensional electron gas (2DEG) is primarily confined to the two or three layers adjacent to the gate dielectric[23,24]. Consequently, the extensive upper layers act as a bulk-like region, which can induce significant electrostatic screening of the applied displacement field and trigger complex interlayer scattering (also Supplementary Note 5). To mitigate these bulk-like interferences and unambiguously isolate the SVL coupling effect, we investigated Device #2 with a reduced thickness of $N_L$ = 5. In this thinner device, the expected consistent behavior was successfully recovered across both energy scales.

**Spin splitting of subbands in Device #2**

The upper panel of Fig. 4a illustrates $G$ as a function of $V_{QPC2}$ for Device #2 at $V_{SD}$ = 2 mV, $V_{BG}$ = 75 V and $V_{QPC1}$ = 4 V across a range of $T$s. In this configuration, the QPC1 channel remains in a highly conductive state (not shown). At $T$ = 30 K, a monotonic increase in conductance is observed for $V_{QPC2} > -4.3$ V. As $T$ decreases, the conductance curve evolves into an oscillatory profile, accompanied by a suppression of the overall conductance magnitude. We attribute this behavior to the electronic states within the QPC channel defined by $V_{QPC2}$. For Device #2, we extracted several key physical parameters: $\lambda_F$ ~ 15.1 nm, $l_e$ ~ 16.4 nm, $W_{eff}$ ~ 5.3 nm and $L_{eff}$ = 10.6 – 26.5 nm channel (see Supplementary Note 3). These values were derived based on a 1D subband confinement energy $\Delta_{qpc}$ ~ 30 meV and an intrinsic mobility $\mu$ ~ 600 $cm^2V^{-1}s^{-1}$, resulting in the relationships $W_{eff} \approx \lambda_F/2$ and $L_{eff} \approx l_e$. This length-scale hierarchy indicates that electrons traversing the QPC constriction have a non-negligible probability of undergoing one or more impurity scattering events before exiting the channel. Because these scattering events inevitably reduce the forward transmission probability below unity ($Tr < 1$), the transport mechanism of Device #2 is unambiguously classified as operating in the quasi-ballistic regime.

To investigate the dependence of $G$ modulation on $V_{BG}$, the lower panel of Fig. 4a presents $G$-$V_{QPC2}$ characteristics for $V_{BG}$ ranging from 70 V to 80 V in 1 V increments at $T$ = 3 K. The curve at $V_{BG}$ = 75 V (highlighted in green) directly corresponds to the data shown in the upper panel of Fig. 4a. At $V_{BG}$ = 80 V, three prominent conductance humps, indicated by green with orange, grey with blue, and purple arrows, define two primary energy scales: $\Delta_1$ and $\Delta_2$. Notably, as $V_{BG}$ decreases, the magnitudes of both $\Delta_1$ and $\Delta_2$ diminish. Simultaneously, two additional energy scales, $\Delta_{1/2}$ and $\Delta_{3/2}$, emerge and progressively increase in magnitude. This systematic evolution is further corroborated by the 2D map of $G$ as a function of $V_{QPC2}$ and $V_{BG}$ in Supplementary Fig. 9, which clearly illustrates the $V_{BG}$ dependence of all four energy scales, despite the leftmost conductance peak for $\Delta_{1/2}$ being partially obscured.

Figures 4b-g display transconductance ($dG/dV_{QPC2}$) maps as a function of $V_{SD}$ and $V_{QPC2}$, with $V_{BG}$ values ranging from 70 V to 80 V in 2 V increments at $T$ = 3 K. At $V_{BG}$ = 80 V (Fig. 4g), the green and blue diamond-shaped features correspond to subband energy intervals of

~12 meV ($\Delta_1$) and ~11 meV ($\Delta_2$), respectively. The colored arrows (orange, green, grey, blue, and purple) in Fig. 4b-g align with the corresponding upward arrows in the lower panel of Fig. 4a. We attribute $\Delta_1$ and $\Delta_2$ to the $G_0$ and $2G_0$ conductance plateaus, respectively.

Notably, the conductance within the first plateau region significantly decreases as temperature falls, as indicated by the dashed oval in the upper panel of Fig. 4a, eventually evolving into a conductance valley. This behavior contrasts with the bulk current increase at lower $T$ observed at $V_{BG}$ = 75 V (see Fig. 1d), confirming that the conductance valley originates specifically from the QPC region rather than the bulk. The $T$-dependent conductance within this dashed oval in the upper panel of Fig. 4a and the accompanying Arrhenius analysis (Fig. 2e, lower panel) provide further evidence for its phase-coherent quasi-ballistic nature. Near the pinch-off regime, conductance is strongly suppressed at low $T$s, and pronounced peak-dip fluctuations emerge below ~ 13 K. The Arrhenius fit in this regime yields an activation energy of $E_a \approx$ 0.9 meV, and tis corresponding thermal energy matches the temperature at which the peak-dip behavior emerges. Rather than indicating a transition to a fully localized quantum dot or a Coulomb blockade regime, this threshold $T$ signifies the onset of phase coherence. Below this $T$, $l_\phi$ expands significantly, and electrons traversing the short saddle-point barrier ($L_{eff} \approx l_e \approx \lambda_F$) undergo coherent resonant backscattering induced by local potential fluctuations (see Supplementary Table 4). This quantum interference reduces the transmission probability below unity ($Tr < 1$), manifesting as the observed resonant peaks.

As $V_{BG}$ decreases, the dimensions of the $\Delta_1$ and $\Delta_2$ diamond structures contract, while two new features, represented by the orange and grey diamonds in Fig. 4f, distinctly emerge at $V_{BG}$ = 78 V. These features evolve into the additional diamond structures $\Delta_{1/2}$ and $\Delta_{3/2}$ shown in Fig. 4e, which expand in size as $V_{BG}$ is further reduced. An analysis of $\delta V_{QPC2}$ as a function of the vertical displacement field ($D$) within the QPC channel gap for $\Delta_n$ ($n = 1 -$

4) reveals that $\Delta_1$ and $\Delta_2$ decrease, whereas $\Delta_{1/2}$ and $\Delta_{3/2}$ increase as $D$ (or $V_{BG}$) is reduced (Fig. 4h). In our specific device architecture, the metallic QPC gates primarily define the lateral confinement potential for the 1D channel. Within the 180 nm gap between the split gates where electron transport occurs, the vertical $D$ is predominantly controlled by the back-gate voltage. The effective $D$ is calculated using the expression: $D = \frac{V_{\mathrm{BG}} - V_{\mathrm{BG}}^{\mathrm{th}}}{t_{\mathrm{SiO_2}}/\varepsilon_{\mathrm{SiO_2}} + t_{h-\mathrm{BN}}/\varepsilon_{h-\mathrm{BN}}}$, where the dielectric constants are $\varepsilon_{h-\mathrm{BN}}$= 4 and $\varepsilon_{\mathrm{SiO_2}}$= 3.9, the dielectric thicknesses are $t_{h-\mathrm{BN}}$= 30 nm and $t_{\mathrm{SiO_2}}$= 500 nm. The back gate threshold voltage is $V_{\mathrm{BG}}^{\mathrm{th}}$ = 34 V. The energy scales for $\Delta_n$ correspond to the diamond dimensions along the $V_{SD}$ axis in Figs. 4b-g, indicated by colored circles that match the respective diamonds in Fig. 4b. Figure 4i shows $\delta V_{SD}^{+}$ as a function of $D$ for $\Delta_1$, $\Delta_2$, $\Delta_{1/2}$, and $\Delta_{3/2}$; across all four subbands, the trends are consistent with those observed in the $\delta V_{\mathrm{QPC2}}$ results in Fig. 4h (see Supplementary Fig. 4b). Here, $\delta V_{SD}^{+}$ is extracted from the positive polarity region (see Supplementary Fig. 6b for $\delta V_{SD}^{-}$). Contrary to the $\Delta_1$ behavior observed for Device #1, Device #2 demonstrates highly consistent trends in $\Delta_n$ across both the $\delta V_{\mathrm{QPC2}}$ and $\delta V_{SD}^{\pm}$ scales. These results indicate that a modest change in the displacement field of ~0.08 V/nm induces a significant spin splitting of approximately 7 meV between opposite spin states, as evidenced by the $\Delta_{1/2}$ and $\Delta_{3/2}$ cases.

**Mechanism of spin-valley-layer coupling**

To interpret the systematic behavior of the four observed subbands under varying electric fields, we consider the electric-field-induced SVL coupling effect[11,12]. Layer polarization is known to occur when the spin-orbit coupling strength exceeds the interlayer hopping amplitude[25], resulting in effective layer-localized, spin-polarized states, a hallmark of SVL coupling. Such localization can lead to layer-specific valley states that are highly tunable via an external electric field. This layer-specific control is further facilitated by the vertical confinement of carriers; specifically, the formation of a 2DEG primarily within the bottom two to three layers of the $WSe_2$ adjacent to the gate dielectric[23,24] may enhance the contrast between valley-polarized states in neighboring layers. Here, we assume that electrons in Q and Q' valleys are predominantly located in the *n*th and (*n*+1)th layers, respectively, adjacent to the gate-dielectric insulator. As illustrated in Fig. 5, the solid purple and red bands represent occupied states for spin-up (↑) and spin-down (↓) orientations in the Q and Q' valleys, respectively. The dashed bands correspond to the upper subbands for electrons confined within the harmonic potential of the QPC structure. An electric field ($E$) is applied perpendicular to the layers via a back-gate voltage.

In the initial state at $E = E_0$ (for a specific $V_{BG}$), as illustrated in Fig. 5a, we assume that the lowest subband in the Q valley (solid purple curve) aligns with the second-lowest subband in the Q' valley (solid purple curve), where both subbands possess the same spin orientation. This alignment yields two subband energy scales, $\Delta_{1/2}$ and $\Delta_{3/2}$. Upon varying $V_{BG}$ such that the electric field increases to $E = E_1$ ($> E_0$) (Fig. 5b), the Q and Q' valleys residing in adjacent layers undergo opposite energy shifts due to the SVL coupling effect, as indicated by the vertical blue arrows. This shift results in the emergence of four distinct subband energy scales: $\Delta_{1/2}$, $\Delta_1$, $\Delta_{3/2}$ and $\Delta_2$. At a higher field strength, $E = E_2$ ($> E_1$) (Fig. 5c), bands with opposite spin orientations from the Q and Q' valleys overlap, simplifying the spectrum to two energy scales of $\Delta_1$ and $\Delta_2$. Specifically, within the investigated electric-field range, $\Delta_{1/2}$ and $\Delta_{3/2}$ are expected to decrease, while $\Delta_1$ and $\Delta_2$ increase with rising field strength.

QPC spectroscopy enables the extraction of energy intervals, denoted as $\Delta_n$, between the subbands formed within the constriction by measuring the differential conductance as a function of $V_{QPC}$ and $V_{SD}$ across a range of $V_{BG}$ values. In Device #2, both $\Delta_1$ and $\Delta_2$ exhibit a systematic increase with $V_{BG}$, whereas $\Delta_{1/2}$ and $\Delta_{3/2}$ decrease as $V_{BG}$ is raised (Fig. 4h,i). This observed evolution aligns perfectly with the proposed mechanism of electric-field-induced spin splitting, as schematically illustrated in the band diagrams of Fig. 5a–c.

The behavior of $\Delta_{1/2}$ and $\Delta_1$ for Device #1, as shown in Fig. 3, represents an intermediate state between the regimes depicted in Fig. 5b, c. As the electric field strength increases, the energy levels of the Q and Q' valleys, localized in distinct layers, undergo downward and upward shifts, respectively. Consequently, the energy scales $\Delta_1$ and $\Delta_{1/2}$ exhibit opposing trends with increasing field strength, a behavior clearly observed along the $\delta V_{QPC3}$ scale in Fig. 3b. However, it should be noted that the subband energy evaluation on the $\delta V_{SD}^{\pm}$ scale for Device #1 is partially inconsistent. This deviation is attributed to the relatively large thickness of Device #1 ($N_L$ = 14), which likely introduces complex interlayer scattering, parasitic states, and screening effects that obscure the idealized subband splitting (see Supplementary Note 5), as aforementioned.

To unambiguously verify the SVL coupling mechanism, our conclusion heavily relies on the results obtained from the thinner Device #2 ($N_L$ = 5). In this device, where bulk-like interlayer effects are minimized, all four subband energy scales ($\Delta_{1/2}$, $\Delta_1$, $\Delta_{3/2}$ and $\Delta_2$) demonstrate perfectly consistent back-gate dependence in both the $\delta V_{QPC2}$ scale and the energy-resolved $\delta V_{SD}^{\pm}$ scale. Furthermore, we clarify that due to the specific device architecture, the QPC split-gates primarily define the lateral confinement potential and serve as a spectroscopic probe, while the vertical $D$ within the QPC channel is predominantly governed by the back gate. Therefore, the highly consistent $\delta V_{SD}^{\pm}$ scaling observed in Device #2 provides robust and direct evidence supporting the electric-field-driven SVL coupling mechanism. This mechanism is further corroborated by Supplementary Fig. 10a-d, which present $dI/dV_{QPC2}$ maps at $V_{BG}$ = 70, 80, 90 and 130 V for Device #2. These measurements, performed in an independent cryogenic system at $T$ = 4 K, reveal a repeated conductance pattern as $V_{BG}$ is increased. This periodicity is entirely consistent with the predicted opposite band shifts in Q and Q' valleys as a function of the applied vertical field, as shown in Supplementary Fig. 10e-h.

**Valley and spin Zeeman effects**

Figure 6a presents $G$ maps as a function of $V_{SD}$ and $V_{QPC2}$ at $B$ = 0 (left panel) and 9 T (right panel) at $V_{BG}$ = 80 V. Figure 6b presents $G$–$V_{SD}$ curves for various $B$ fields at the $V_{QPC2}$ positions corresponding to the $\Delta_{1/2}$ region (indicated by white arrows in Fig. 6a), where the $G$–$V_{QPC2}$ curve at zero bias exhibits a local conductance maximum (see Supplementary Figure 11). At $B$ = 0 T, two local conductance peaks are observed near $V_{SD}$ = $\pm$3 mV (indicated by green triangles), with a conductance dip at zero bias. As $B$ increases, the peak separation widens and the conductance dip deepens. The upper panel of Fig. 6d shows $\Delta_{1/2}$ (= the peak interval, $e\delta V_{\mathrm{SD}}$) as a function of $B$, fitted by the relation $e\delta V_{\mathrm{SD}} = |g|\mu_B B$ $(= 2\Delta E_z)$, yielding $|g|$ = 3.45 $\pm$ 0.85 for electrons, where $g$ is the Landé $g$-factor and $\mu_B$ is the Bohr magneton. Here, $\Delta E_z$ is the Zeeman energy shift. Meanwhile, Fig. 6c shows $G$–$V_{SD}$ curves for various $B$ fields obtained at the locations indicated by black arrows in Fig. 6a, corresponding to the $\Delta_1$ region, where the $G$–$V_{QPC2}$ curve at zero bias exhibits a local conductance minimum (see Supplementary Figure 11). Experimental observations indicate that the conductance peaks, indicated by green arrows in Fig. 6c, relatively insensitive to variations in the $B$ field up to 9 T. The lower panel of Fig. 6d shows $\Delta_1$ as a function of $B$, where the dashed red fit yields $|g| \approx$ 0.52 $\pm$ 0.5. We suggest that the obtained $|g|$ values for $\Delta_{1/2}$ and $\Delta_1$ regions, respectively, are attributed to the combined contributions of spin ($g_s = 2$) and valley ($g_v$) components. This is described by the relation of $g = s_z g_s \pm \kappa g_v$, where $s_z$ = 1 ($-1$) represents an up(down)-spin electron and $\kappa$ = 1($-1$) denotes the Q (Q') valley[9].

The left panel of Fig. 6e shows the bulk case where $g_v$ = 4[7,26], resulting in $g$ = 2, −2, 6, and −6 for $Q_{\downarrow}$, $Q'_{\uparrow}$ $Q_{\uparrow}$, and $Q'_{\downarrow}$ states, respectively. These values correspond to Zeeman energy shifts of $\Delta E_z = \mu_B B$, $-\mu_B B$, $3\mu_B B$, and $-3\mu_B B$, respectively, relative to the top subband. Conversely, the right panel of Fig. 6e depicts the case where $g_v$ = 0. Based on our experimentally observed $|g|$ values, we consider the case where $g_v = 1$, as shown in the middle panel of Fig. 6e. Under this condition, we obtain $g$ = −1, 1, 3, and −3 for $Q_{\downarrow}$, $Q'_{\uparrow}$ $Q_{\uparrow}$, and $Q'_{\downarrow}$, respectively. Here, $\Delta_{1/2}$ increases with the $B$ field, while $\Delta_1$ gradually decreases, consistent with the trends observed in Fig. 6d.

We propose that the significant reduction in the $g_v$ from its bulk value originates from the partial quenching of the orbital magnetic moment. This quenching is likely induced by the strong lateral confinement within the QPC subband. Since the valley Zeeman effect originates from the orbital magnetic moment associated with the Berry curvature, the spatial restriction of the electronic wave packet physically suppresses the orbital contribution[27]. This orbital phenomenon is sensitive to spatial confinement; in narrow nanostructures, the orbital angular momentum can undergo significant quenching[28]. Magneto-optical experiments indicate that suggest that a robust valley $g$-factor requires a minimum spatial extent, typically a radius of 2 to 4 nm[29], to sustain the circulating current loops associated with the valley degree of freedom. In our device, the ground-state subband is confined to a effective width of approximately 5 nm. This regime of hard confinement physically constricts the electronic wave packet, thereby suppressing the orbital contribution to the total Zeeman energy.

Furthermore, we highlight the exceptional efficiency of electrical control in this system. A variation in $V_{BG}$ of 10 V, corresponding to a displacement field change of $\Delta D$ ~ 0.08 V/nm, induces a spin splitting of ~7 meV. In contrast, a magnetic field of $B$ = 9 T yields a Zeeman splitting of only ~2 meV. This result demonstrates that the electric-field-induced SVL coupling provides a significantly more powerful means of manipulating spin and valley degrees of freedom than conventional magnetic field approaches.

In summary, we have demonstrated robust, electric-field-controlled spin splitting within the Q and Q' valleys of multilayer $WSe_2$ using QPC spectroscopy. Our results show that an out-of-plane gate voltage systematically modulates four distinct energy scales ($\Delta_n$) extracted from transconductance maps. This provides direct spectroscopic evidence of the dynamic evolution of the spin-resolved density of states within the QPC constriction. Importantly, the magnitude of this electric-field-induced splitting, ~7 meV for a displacement field change of ~0.08 V/nm, far exceeds the ~2 meV splitting achievable by a 9 T magnetic field, highlighting the exceptional tunability of spin-orbit coupling in the conduction band of multilayer $WSe_2$. We attribute this behavior to the electric-field-induced SVL coupling mechanism, wherein the gate voltage modulates the energy alignment of Q and Q' valleys across adjacent layers. This interpretation is directly supported by the observation of subband energy scales that exhibit contrasting and systematic dependencies on the back-gate voltage. By circumventing the requirement for high magnetic fields, this SVL-driven control enables the precise manipulation of layer-polarized valley states through purely electrical means. Our findings establish a novel and energy-efficient pathway for engineering spin and valley degrees of freedom, paving the way for next-generation low-power spintronic and quantum information technologies based on TMDC heterostructures.

## Data availability

The datasets generated and analyzed during the current study are available in Zenodo, DOI: 10.5281/zenodo.20350179.

## Acknowledgements

We thank Prof. Sang-Jun Choi and Dr. In-Ho Lee for valuable discussions. This work was supported by the National Research Foundation funded by the Korean Government (RS-2021-NR059826, RS-2023-00283291, RS-2022-NR075640). This research was partially supported by the Korea Research Institute of Standards and Science (KRISS-2026-GP2026-0010) and the National Research Council of Science & Technology (GTL25011-000). K.W. and T.T. acknowledge support from the JSPS KAKENHI (21H05233 and 23H02052), the CREST (JPMJCR24A5), JST and World Premier International Research Center Initiative (WPI), MEXT, Japan.

## Author contributions

J.J.K. and M.H.B. conceived and designed the experiments. K.W. and T.T. grew the bulk *h*-BN. M.G.K. fabricated the devices, performed the measurements, and analyzed the data with help from M.S.K. M.G.K. and M.H.B. wrote the manuscript with contributions from all authors.

## Competing interests

The authors declare no competing interests.

## Additional information

**Supporting Information** The online version contains supplementary material available at …

**Correspondence** and requests for materials should be addressed to Ju-Jin Kim and Myung-Ho Bae.

**Reprints** and permissions information is available at…

# FIGURES

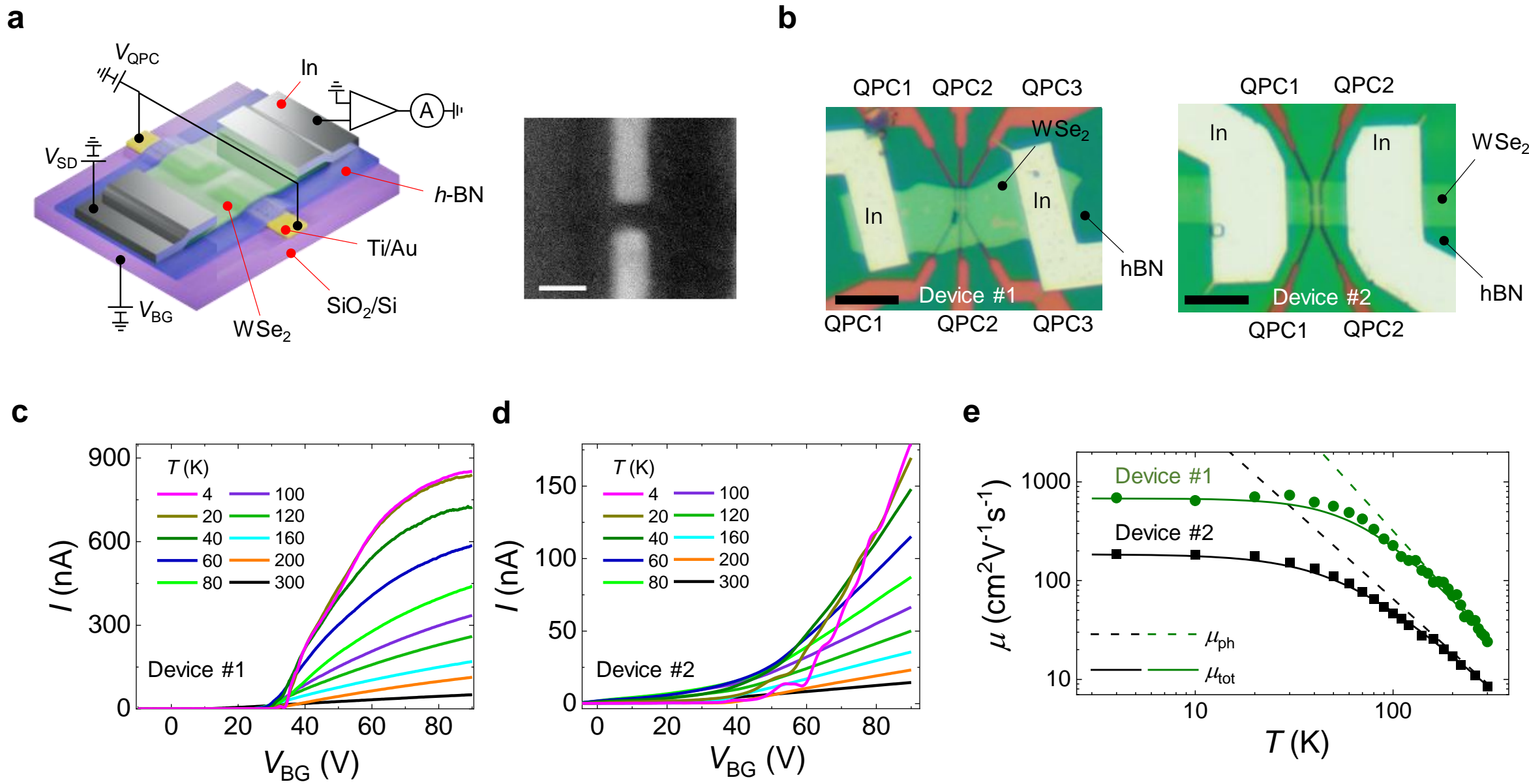

**Fig. 1| Multilayer $WSe_2$ quantum point contact devices and basic characterization. a** Left: Schematic of the $WSe_2$ QPC device and measurement configuration. Right: SEM image of the split metal gates with a QPC channel gap of ~180 nm (scale bar: 300 nm). **b** Optical images of Device #1 (left) and Device #2 (right). Scale bars: 5 μm. **c**,**d** $I$-$V_{BG}$ curves at various temperatures for Device #1 (**c**) and Device #2 (**d**), measured at $V_{SD}$ = 10 and 5 mV, respectively. **e** Extracted $\mu$-$T$ curves for both devices. Solid and dashed curves: fitting lines.

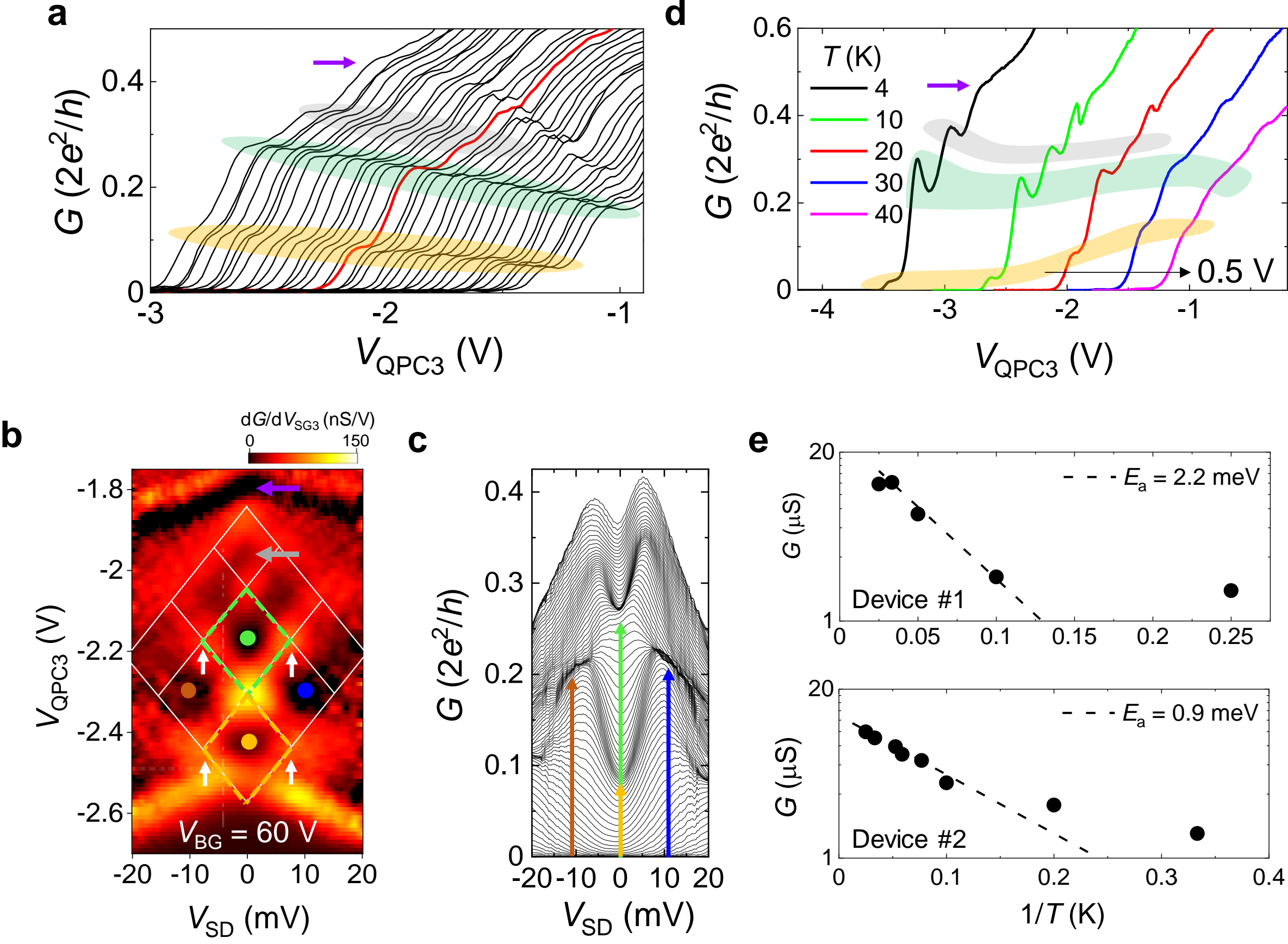

**Fig. 2| Finite-bias subband spectroscopy and *T*-dependent conductance quantization. a** $G$-$V_{QPC3}$ curves of Device #1 at $T$ = 20 K, with $V_{BG}$ values ranging from 68 V (leftmost curve) to 50 V (rightmost curve) in increments of 0.5 V. The orange-, green- and grey-colored regions correspond to conductance-step groups. The red curve represents data obtained at $V_{BG}$ = 60 V. **b** $dG/dV_{QPC3}$ as a function of $V_{SD}$ and $V_{QPC3}$ at $V_{BG}$ = 60 V. The colored features of orange and green, and the region indicated by the grey arrow correspond to the shaded regions of the same colors in (**a**). **c** $G$-$V_{SD}$ curves at various $V_{QPC3}$ values ranging from −1.9 V (top curve) to −2.7 V in increments of 10 mV. **d** $G$-$V_{QPC3}$ curves for various temperatures at $V_{BG}$ = 60 V and $V_{SD}$ =

0. For data obtained at $T = 20$, 30, and 40 K are horizontally offset in increments of 0.5 V, as indicated by the black arrow. The conductance-step groups indicated by the purple arrows in (**a**) and (**d**), along with the dark region marked by the purple arrow in (**b**), originate from a different physical mechanism than the rest groups (see supplementary Fig. 7). **e** Semi-logarithmic plots of $G$ as a function of $1/T$ of Device #1 (upper panel) and Device #2 (lower panel). For Device #1, $G$ was extracted from the first plateau observed near the pinch-off voltage in (**d**). For Device #2, $G$ was extracted from the local minimum $G$ within the dashed oval in the upper panel of Fig. 4**a**. The dashed lines represent the linear fit used to extract the activation energy ($E_a$).

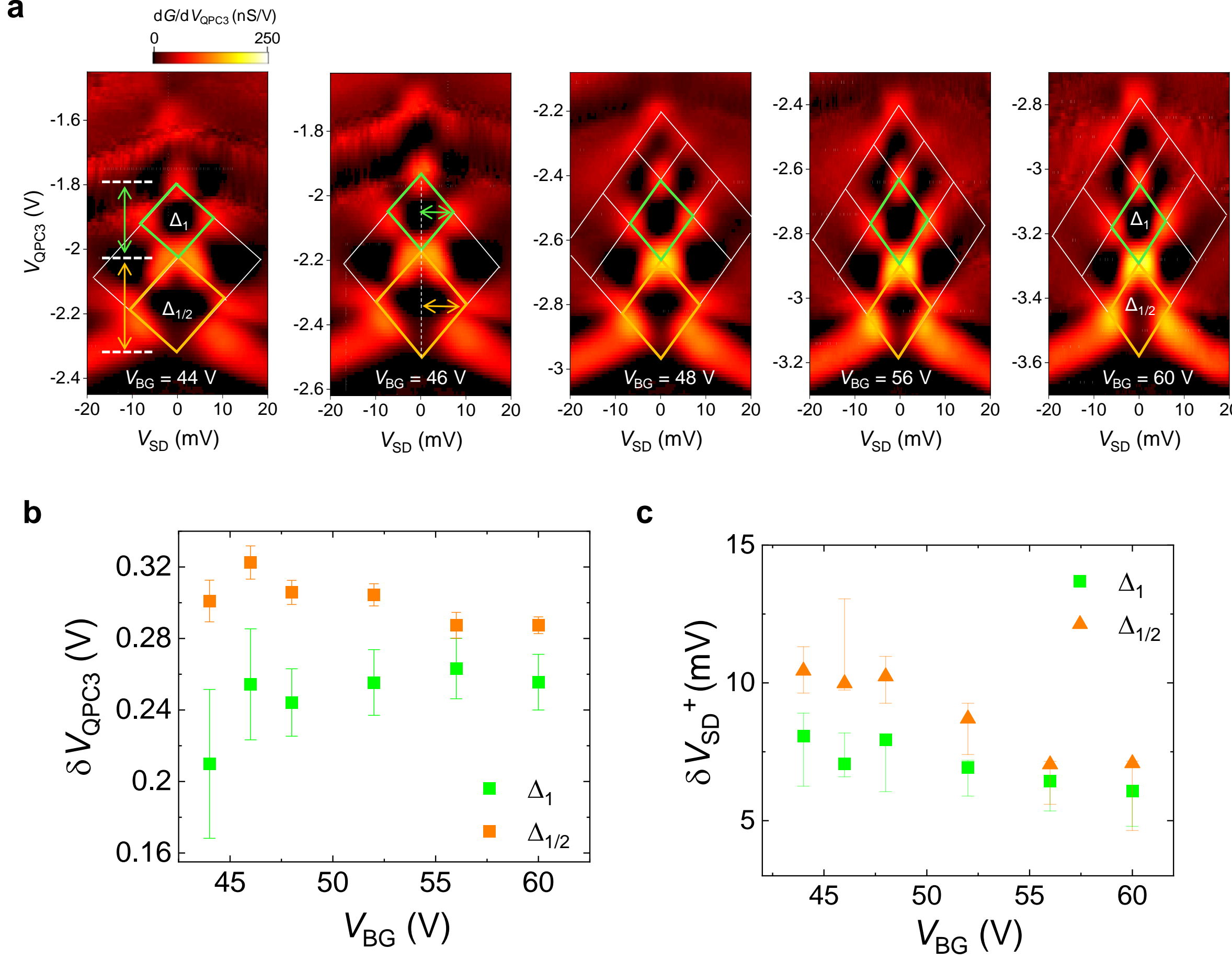

**Fig. 3| Back-gate voltage dependence of the 1D subband energy scales of Device #1. a** $dG/dV_{\mathrm{QPC3}}$ as a function of $V_{\mathrm{SD}}$ and $V_{\mathrm{QPC3}}$, with $V_{\mathrm{BG}}$ values ranging from 44 V to 60 V. The scale bar in the leftmost panel applies to all other plots. **b** Extracted $\delta V_{\mathrm{QPC3}}$ as a function of $V_{\mathrm{BG}}$ for the $\Delta_{1/2}$ and $\Delta_1$ subbands. The $\delta V_{\mathrm{QPC3}}$ values were obtained from the $V_{\mathrm{QPC3}}$ interval defined by the vertical green and orange bidirectional arrows for the corresponding $\Delta_n$ locations. **c** Extracted $\delta V_{SD}^{+}$ as a function of $V_{\mathrm{BG}}$ for $\Delta_{1/2}$ and $\Delta_1$. The $\delta V_{SD}^{+}$ values correspond to the energy scales indicated by the horizontal green and orange bidirectional arrows in the second plot of (**a**) ($V_{\mathrm{BG}} = 46$ V). The error bars in (**b**) and (**c**) are determined based

on the flatness of the $dG/dV_{QPC3}$ peaks within a range of ~3 nS/V, utilizing a noise-bounded error bar evaluation procedure (see Supplementary Note 9).

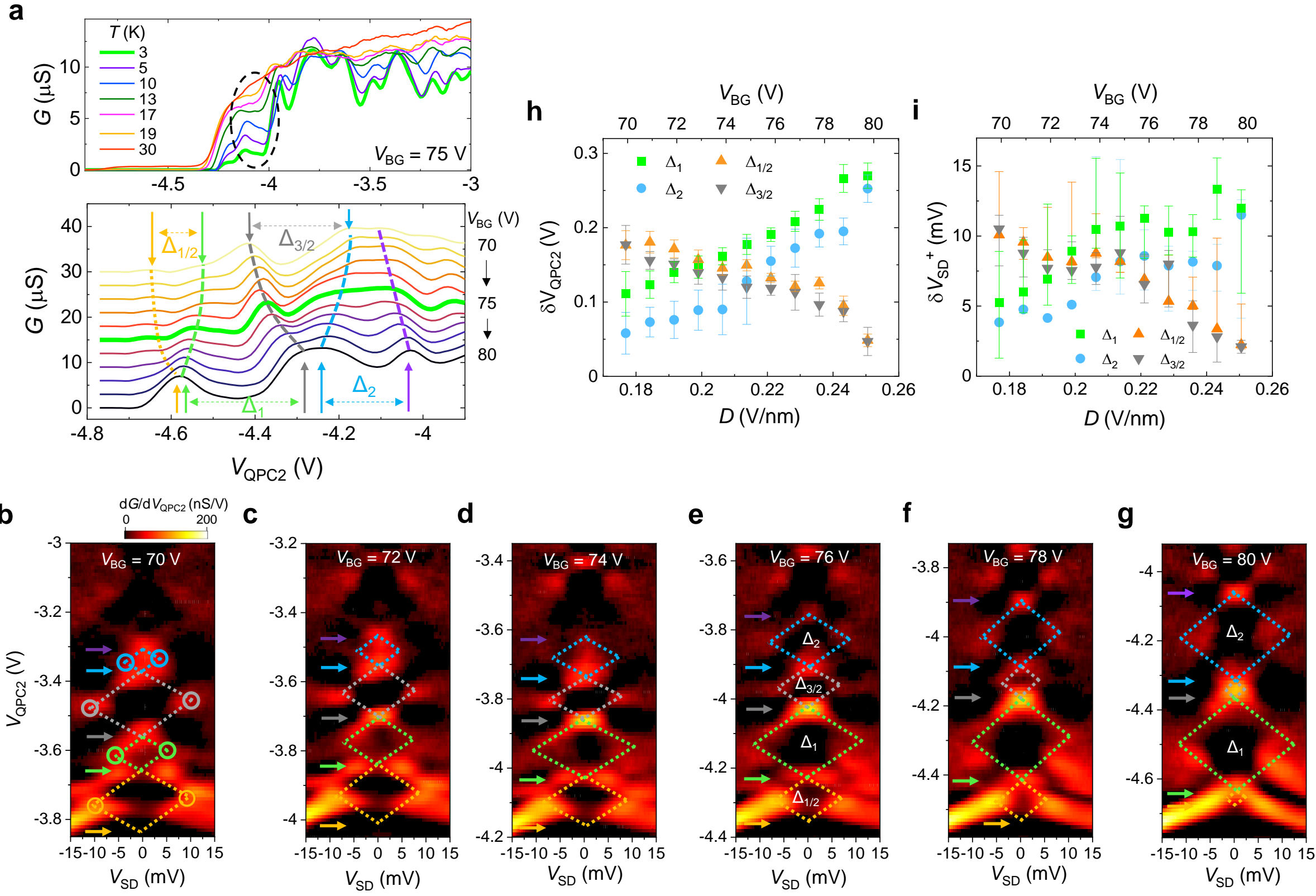

**Fig. 4| Displacement-field dependence of the spin-resolved subbands in Device #2. a** Top panel: $G$ as a function of $V_{QPC2}$ of Device #2, measured at $V_{SD}$ = 2 mV with $V_{BG}$ = 75 V for various $T$s. Bottom panel: $G$-$V_{QPC2}$ curves at $T$ = 3 K for various $V_{BG}$ values ranging from 70 V to 80 V in 1 V increments. The curves are horizontally offset relative to the $V_{BG}$ = 80 V curve to align pinch-off voltages, and vertically offset in increments of 3 μS for clarity. $\Delta_n$ denotes the energy scales of corresponding subbands observed in the QPC structure. **b-g** $dG/dV_{QPC2}$ as a function of $V_{SD}$ and $V_{QPC2}$ for $V_{BG}$ values ranging from 70 V to 80 V in increments of 2 V. The color scale bar in (**b**) applies to all maps in (**c-g**). **h** Extracted $\delta V_{QPC2}$ as a function of $D$ and $V_{BG}$ for the four subbands ($\Delta_n$). **i** Extracted $\delta V_{SD}^{+}$ as a function of $D$ and $V_{BG}$ for $\Delta_n$. The

error bars in (**h**) and (**i**) are determined based on the flatness of the $dG/dV_{QPC2}$ peaks within a range of ~3 nS/V, utilizing a noise-bounded error bar evaluation procedure (see Supplementary Note 9).

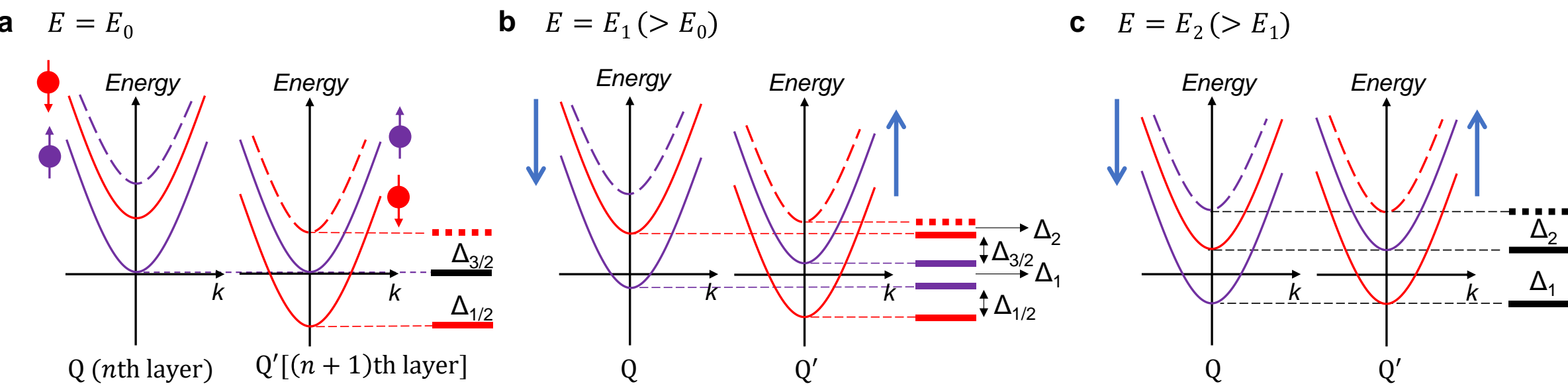

**Fig. 5| Electric-field-induced evolution of spin-resolved subbands via spin-valley-layer coupling. a-c** Schematic illustration of the SVL coupling-induced band shifts in the Q and Q' valleys of the conduction band, which are localized in the *n*-th and (*n*+1)-th layers of multilayer $WSe_2$, respectively. The energy band alignments and the resulting subband energy scales ($\Delta_n$) are shown for varying perpendicular electric fields: **(a)** $E = E_0$, **(b)** $E = E_1$ ($> E_0$), and **(c)** $E = E_2$ ($> E_1$). The vertical blue arrows indicate the opposite energy shifts of the adjacent layers driven by the increasing electric field.

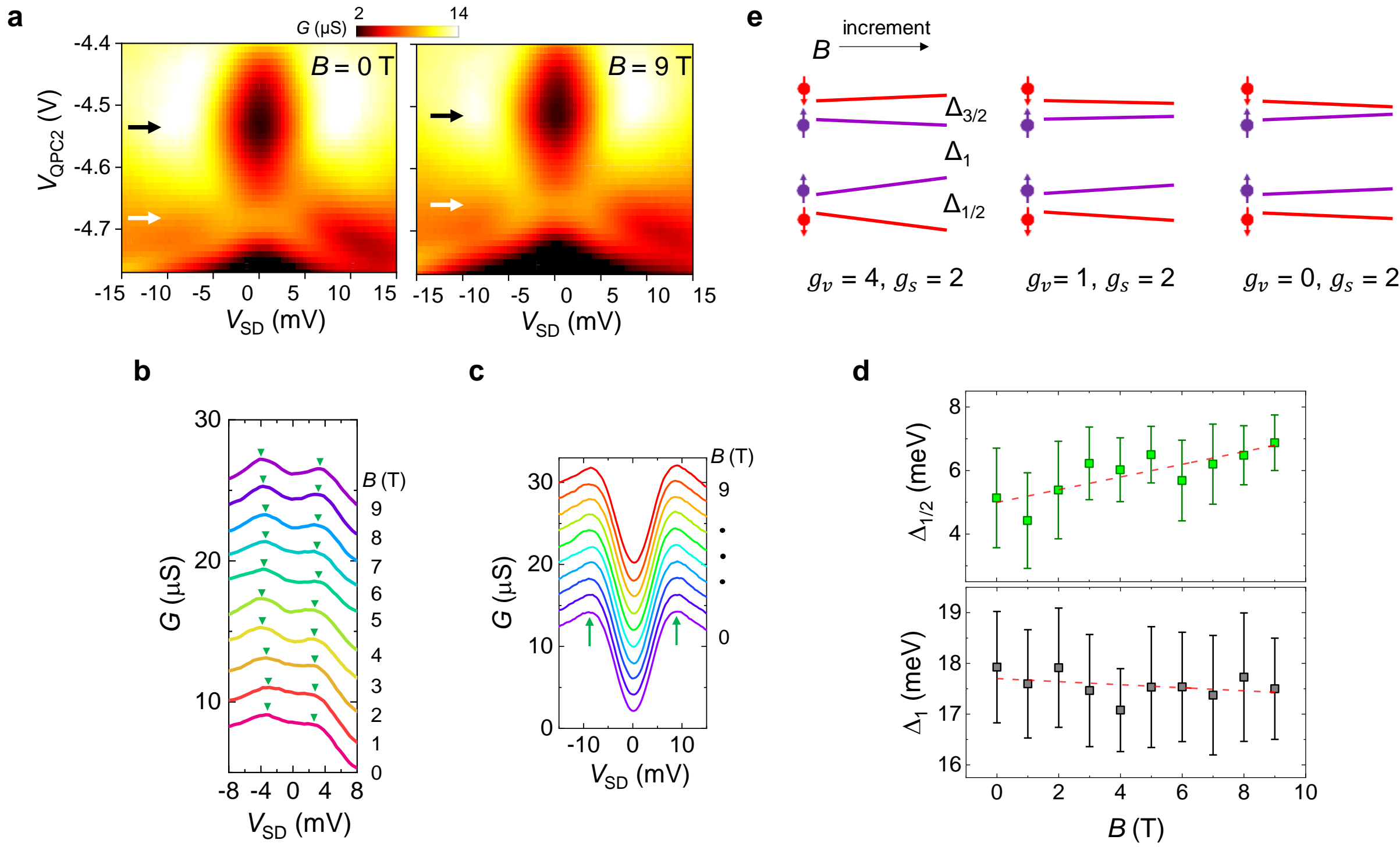

**Fig. 6| Magnetic-field response of the spin-resolved subbands and extraction of the Landé g-factor. a** $G$ as a function of $V_{\mathrm{SD}}$ and $V_{\mathrm{QPC2}}$ at $B$ = 0 (left panel) and 9 T (right panel) at $V_{\mathrm{BG}}$ = 80 V. **b** $G$–$V_{\mathrm{SD}}$ curves for various $B$ fields at the $V_{\mathrm{QPC2}}$ locations indicated by the white arrows in (**a**), corresponding to the $\Delta_{1/2}$ energy scale. **c** $G$–$V_{\mathrm{SD}}$ curves for $B$ = 0 to 9 T in increments of 1 T at the $V_{\mathrm{QPC2}}$ locations indicated by the black arrows in (**a**), corresponding to the $\Delta_1$ energy scale. The curves in (**b**) and (**c**) are vertically offset by 2 μS for clarity. **d** Upper panel: extracted $\Delta_{1/2}$ as a function of $B$, where $\Delta_{1/2}$ is defined as the peak interval ($\delta V_{\mathrm{SD}}$), corresponding to the distance between the two green inverted triangles on each curve in (**b**). The red dashed line represents the linear fit result. Lower panel: extracted $\Delta_1$ as a function of $B$, where $\Delta_1$ is defined by half the peak interval in (**c**), as indicated by the vertical dotted lines. The error bars in (**d**) are determined based on the flatness of the $G$ peaks, utilizing a noise-bounded error bar evaluation procedure. The error bars in (**d**) are determined based on the flatness of the $G$ peaks within a range of ~40 nS and 60 nS for $\Delta_{1/2}$ and $\Delta_1$, respectively, utilizing a noise-bounded error bar evaluation procedure (see Supplementary Note 9). **e**

Schematic of the subband energy evolution when a perpendicular $B$-field is applied to the device, illustrated for three different valley $g$-factor ($g_v$) scenarios.

Supplementary Information for

# Non-magnetic spin splitting driven by spin-valley-layer coupling in multilayer $WSe_2$

Min-Gue Kim[1,2], Min-Sik Kim[1,2], Kenji Watanabe[3], Takashi Taniguchi[4], Ju-Jin Kim[1*], and Myung-Ho Bae[2,5*]

[1]Department of Physics, Jeonbuk National University, Jeonju 54896, Republic of Korea

[2]Korea Research Institute of Standards and Science, Daejeon 34113, Republic of Korea

[3]Research Center for Functional Materials, National Institute for Materials Science, 1-1 Namiki, Tsukuba 305-0044, Japan

5International Center for Materials

[4]Nanoarchitectonics, National Institute for Materials Science, 1-1 Namiki, Tsukuba 305-0044, Japan

[5]KAIST Graduate School of Quantum Science and Technology, Korea Advanced Institute of Science and Technology, Daejeon 34141, Republic of Korea

*e-mail: jujinkim@chonbuk.ac.kr, mhbae@kriss.re.kr

## Supplementary Note 1. Device fabrication and measurement setup

### Device fabrication

A 200 nm-wide Ti(5 nm)/Au(15 nm) split-gate electrode was designed for a QPC configuration on a $SiO_2$(500 nm)/Si substrate, with an interval of 180 nm between the split gates. *h*-BN and $WSe_2$ flakes were then sequentially stacked using mechanical exfoliation and the dry-transfer method using PDMS (polydimethylsiloxane) films, ensuring precise alignment with the split gates. The few tens of nanometer-thick *h*-BN insulating layer serves as the dielectric material between the $WSe_2$ multilayer and QPC gates. The $WSe_2$ layer was in contact with 100 nm-thick In metal, forming the source and drain electrodes. In particular, the In metal was deposited on a substrate holder cooled to 100 K with liquid nitrogen. Supplementary Fig. 1 shows AFM images of fabricated devices, Device #1 and Device #2.

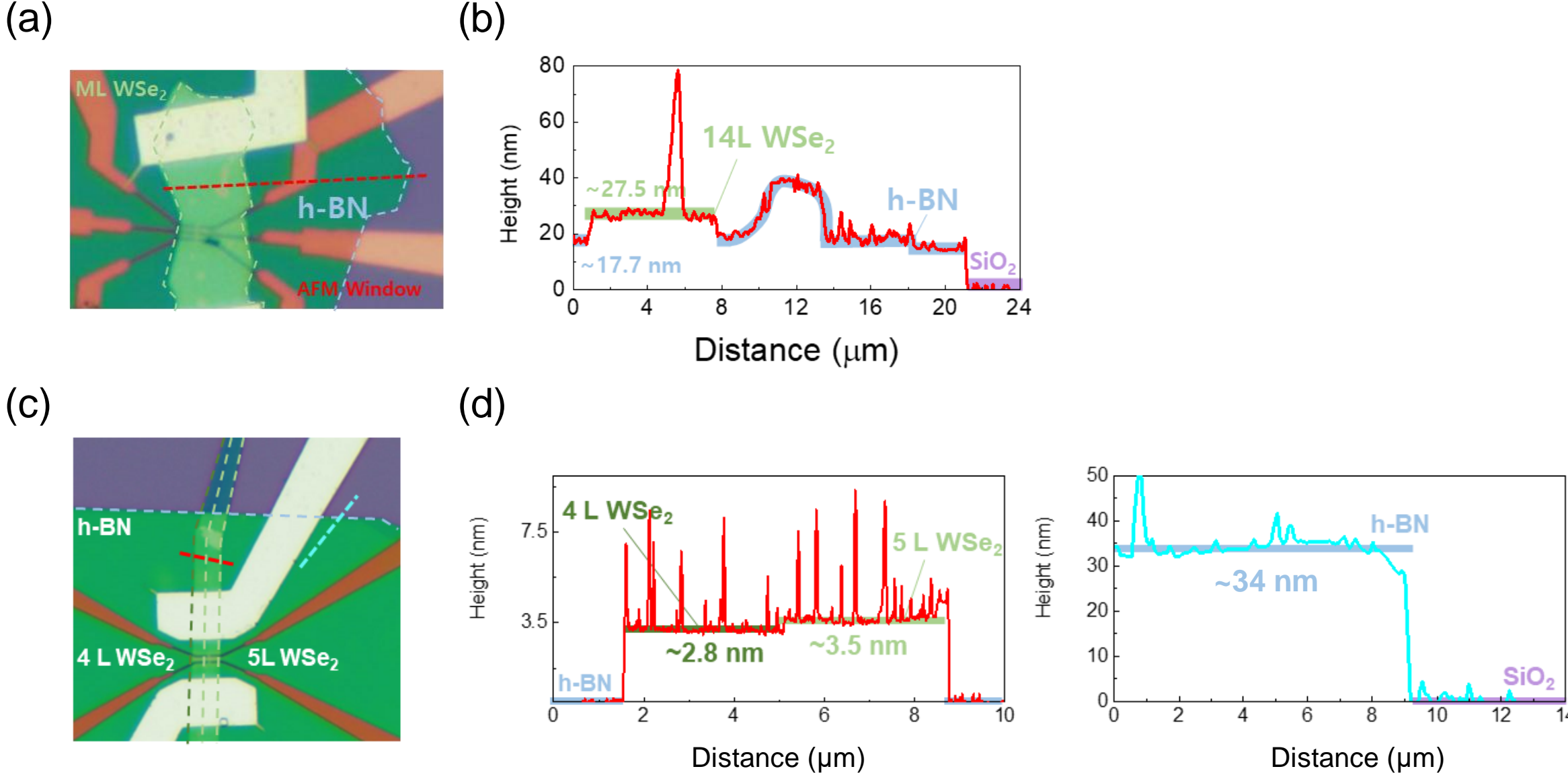

**Supplementary Figure 1.** (a) Optical photograph image of Device #1. (b) AFM height profile along the red dashed line in (a). (c) Optical photograph image of Device #2. (d) AFM height profiles along the red (left panel) and cyan (right panel) dashed lines in (c). Here, the QPC is located in a region having the height of 3.5 nm. If we considering a single layer thickness of $WSe_2$ as 0.7 nm, we estimated the layer numbers of Device #1 and Device #2 as $N_L = 5$ and 14, respectively.

### Measurement setup

Electrical measurements were performed using a two-probe configuration in a cryogen-free system for device#1 and in a physical property measurement system (PPMS) equipped with a 9 T magnet for device#2. Source-drain and gate voltages were applied using Yokogawa 7651 voltage sources. The output current was amplified by a DL 1211 transimpedance amplifier and recorded with a Keithley 2000 voltmeter.

**Supplementary Note 2. Rules for eye-guide lines for subband structures in transconductance maps**

Supplementary Figure 2 displays $dG/dV_{QPC3}$ map as a function of $V_{QPC3}$ and $V_{SD}$ at $V_{BG}$ = 60 V for Device #1. The vertical white line indicates the zero-bias axis. To accurately extract the 1D spatial confinement energy ($\Delta_{QPC}$) and SVL splitting scales ($\Delta_n$) from these maps, the eye-guide lines were constructed based on following physical resonance conditions:

1. All eye-guide lines must originate strictly from the transconductance peaks (nodes) located on the zero-bias axis ($V_{SD}$ = 0 V). These nodes represent the physical resonance conditions where the bottom of each subband exactly aligns with the equilibrium Fermi level.
2. The slope of the transconductance line is given by at least two transconductance peaks. The opposite slopes of transconductance lines follows the same rule [see (i)-(iii) in Fig. Supplementary Fig. 2a].
3. Over a narrow gate voltage range where a few subbands open, the geometric capacitance remains nearly constant. Therefore, the positive slope must be drawn perfectly parallel across all consecutive subbands. The same parallel constraint strictly applies to the negative slope.

Supplementary Fig. 2a, transconductance lines crossing the top vertex denoted by (iv), corresponding a transconductance peak along the zero-bias white line, are justified by the 3rd rule. However, it was not straightforward to draw the transconductance lines crossing the bottom vertex denoted by (v), because there is no transconductance peak at the vertex. On the other hand, in Fig. Supplementary Fig. 2b at $V_{BG}$ = 48 V, two transconductance peaks are found at locations (i) and (ii) along two transconductance lines with a opposite slopes. We note that other transconductance peaks indicated by (iii) and (iv) could mislead in the finding right

transconductance peaks to draw the transconductance lines. Considering this situation, the location (vi) in Fig. Supplementary Fig. 2a was found. Then, with the rules of 1-3, we draw the corresponding opposite line, resulting in a vertex at the location (v). Figure Supplementary Fig. 3 and Supplementary Fig. 4 show the transconductance lines on transconductance maps obtained for all examined $V_{BG}$ conditions of Device #1 and Device #2, respectively.

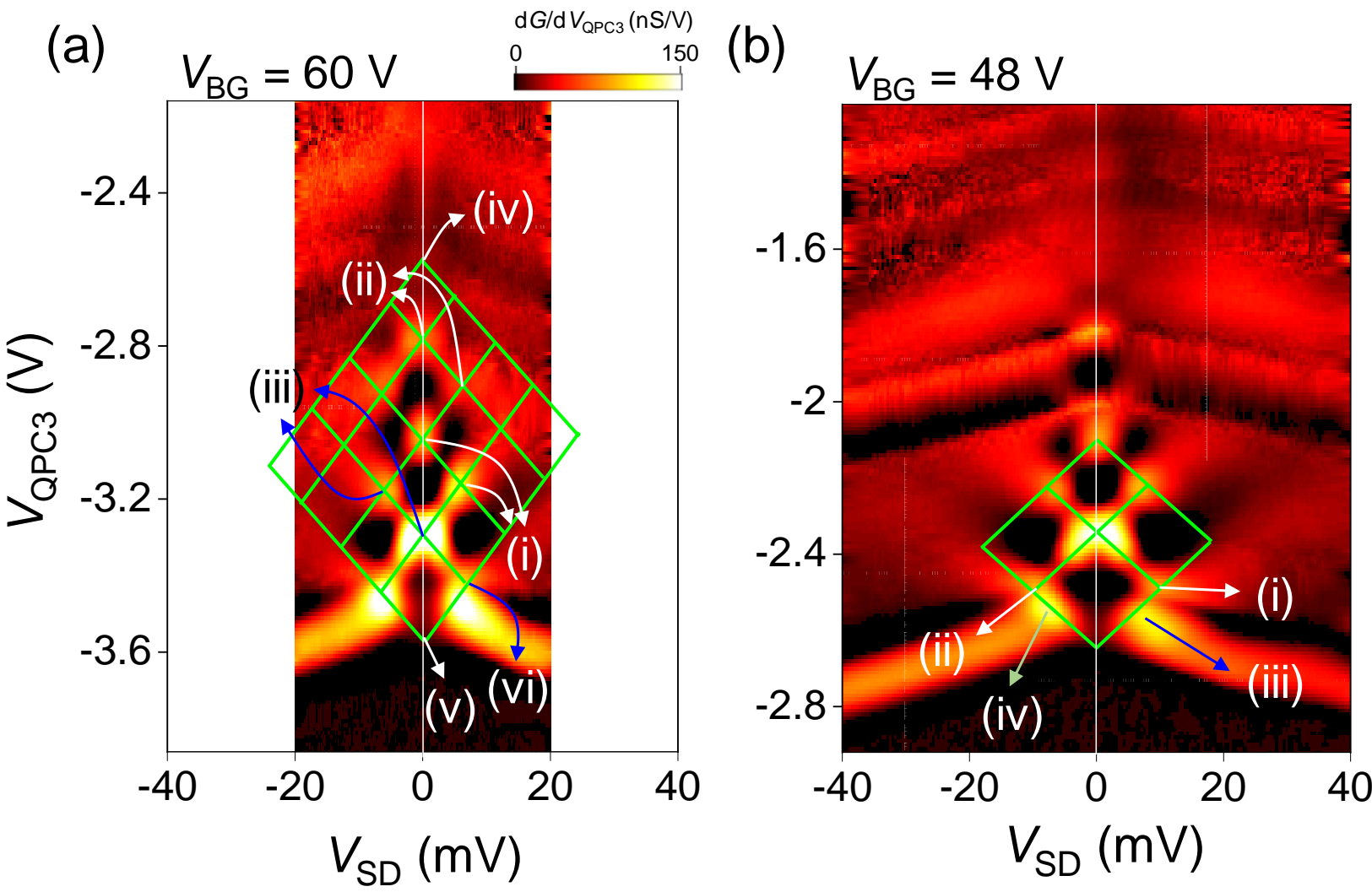

**Supplementary Figure 2.** Extended data of Fig. 3a at (a) $V_{BG}$ = 60 V and (b) 48 V of Device #1.

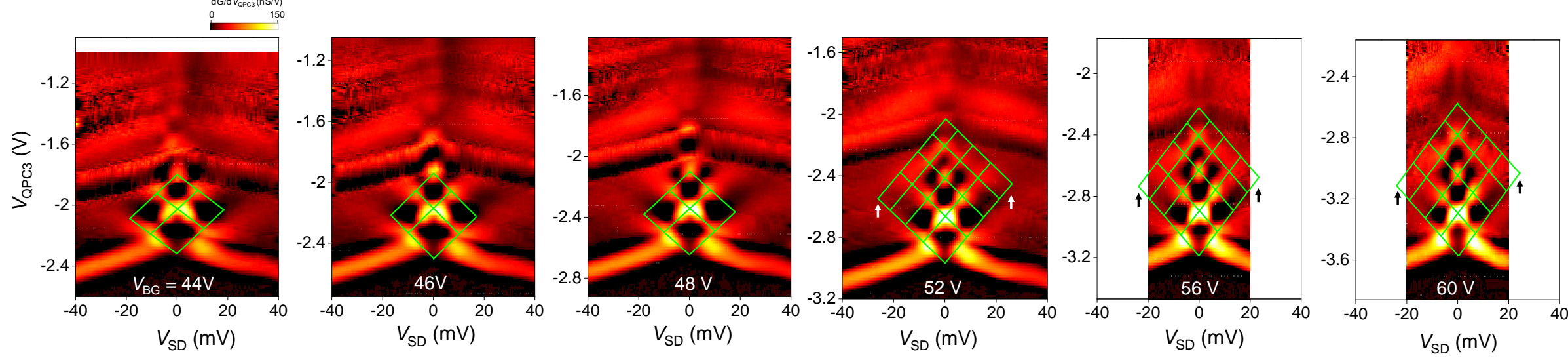

**Supplementary Figure 3.** Extended data of Fig. 3a (Device #1), including data obtained at $V_{BG}$ = 52 V.

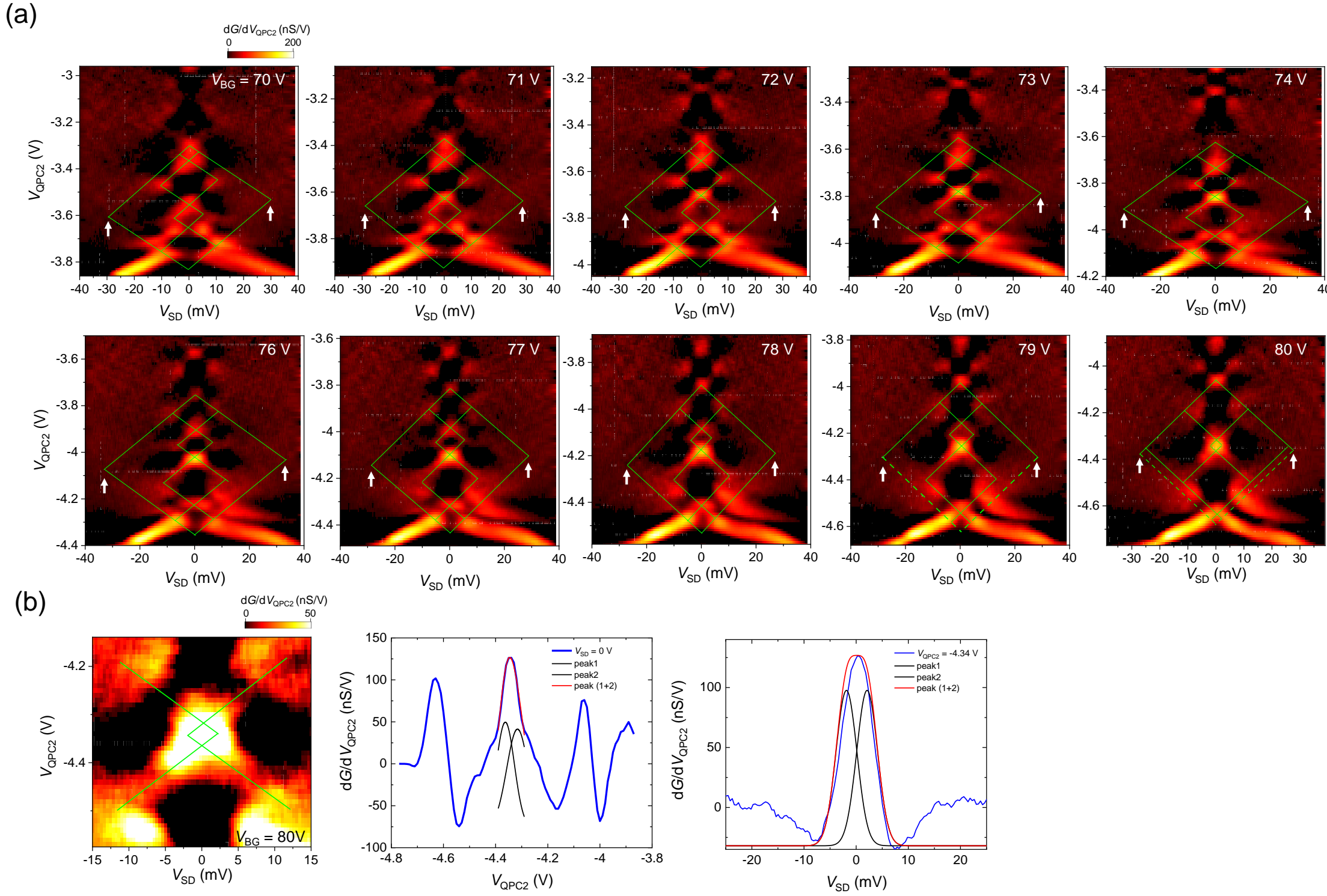

**Supplementary Figure 4.** (a) Extended data from data in Fig. 4b-g (Device #2), including data obtained at $V_{BG} = 71$, 73, 77, and 79 V. (b) Left: zoom-in plot of 2D map at $V_{BG} = 80$ V. Middle and right: fits to extract subband energy scales.

## Supplementary Note 3. Length Scales and transport regimes of devices

To quantitatively evaluate the electrostatic landscape of our QPC, we deduce the effective channel dimensions based on finite-bias spectroscopy measurements and the Büttiker saddle-point potential model[22], utilizing the intrinsic 1D subband energy gaps ($\Delta_{\mathrm{QPC}}$) extracted directly from our finite-bias transconductance measurements (accounting for SVL coupling). It turns out that the transport regimes of both Device #1 and Device #2 are not ballistic regimes, but quasi-ballistic regimes with a transmission probability below unity ($Tr < 1$).

### 1D spatial confinement energy

To accurately extract the intrinsic 1D spatial confinement energy (pure QPC subband energy, $\Delta_{\mathrm{QPC}}$) from the transconductance spectroscopy, it is essential to account for the spin-valley-layer (SVL) coupling induced by the perpendicular electric field. Because the SVL coupling significantly lifts the degeneracy of the bands, a single fundamental 1D subband is split into multiple spin-resolved levels. As illustrated in the band diagram in Supplementary Fig. 5d under strong SVL coupling, the fundamental 1D subband splits into four distinct spin-resolved energy levels. Consequently, in the corresponding finite-bias transconductance maps (Supplementary Fig. 5a for Device #1 at $V_{BG}$ = 60 V and Figure Supplementary Fig. 5b, c for Device #2 at $V_{BG}$ = 76 V and $V_{BG}$ = 80 V, respectively), the pure 1D subband energy is determined by the total energy span that encompasses all four nested split-diamonds, indicated by the large outer green diamond structures. Supplementary Fig. 5d, e show the subband structure with relatively lower and higher $V_{BG}$ values, respectively. For instance, Supplementary Fig. 5a and b correspond to Supplementary Fig. 5d, and Fig. Supplementary Fig. 5c is a region near the condition of Fig. Supplementary Fig. 5e.

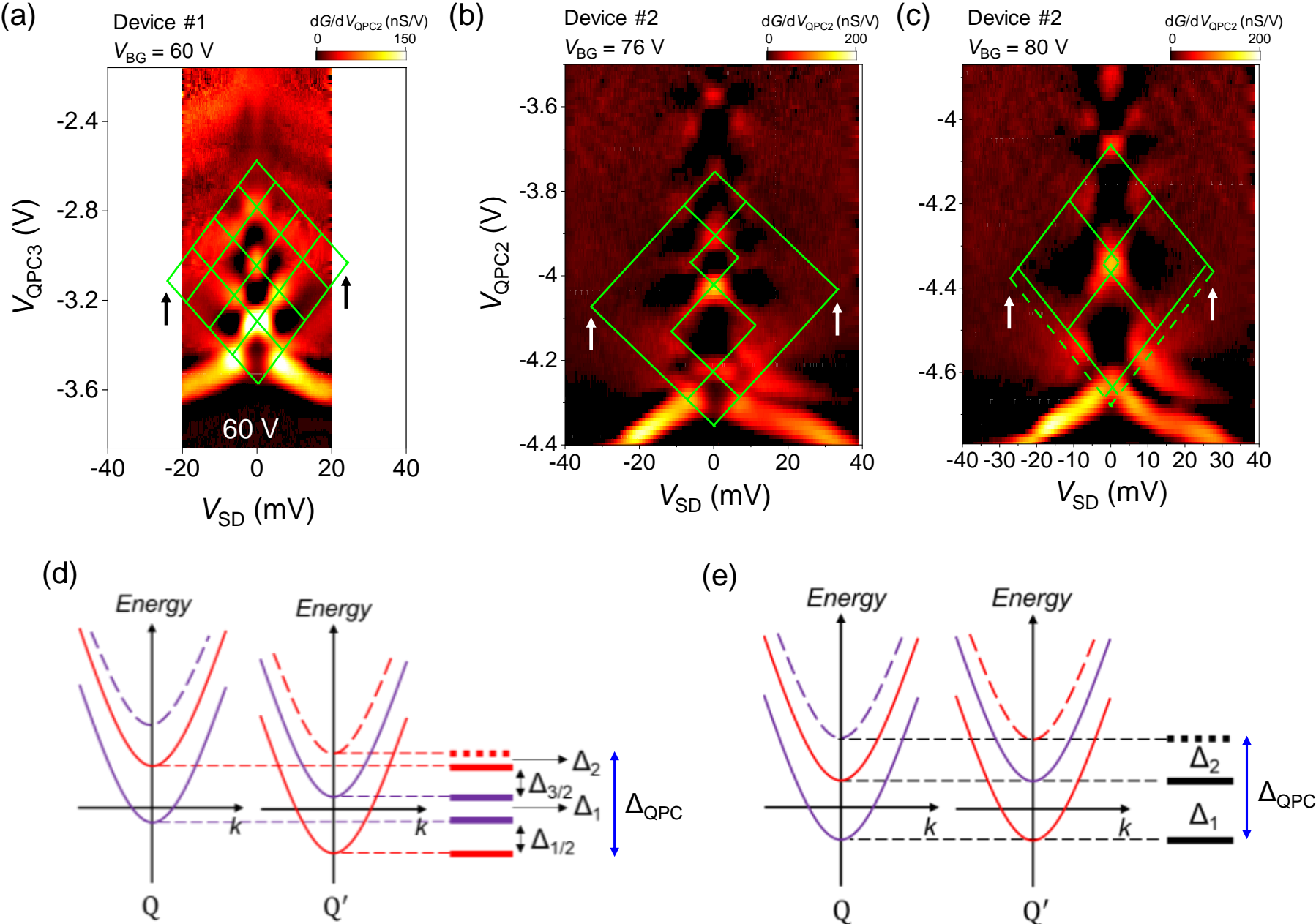

**Supplementary Figure 5.** (a)-(c) Extension versions of Fig. 3a ($V_{BG}$ = 60 V, Device #1), Figs. 4e, g ($V_{BG}$ = 76 and 80 V, Device #2), respectively, to extract 1D spatial confinement energy scales of Device#1 and Device#2. (d),(e) Subband splitting based on the SVL coupling model. Left and Right: relatively lower and higher $V_{BG}$ values, respectively.

By systematically applying this SVL-inclusive geometric analysis to the transconductance maps, we can reconstruct the pure electrostatic confinement energies. For the cases of Fig. Supplementary Fig. 5a-c, we find that $\Delta_{QPC} \approx$ 24 meV for Device #1 and 33.2 and 27.5 meV for Device #2, respectively, based on the half value of the interval of $V_{SD}$ values ($\Delta V_{SD}/2$) indicated by two arrows. By examined the $dG/dV_{QPC}$ maps across various $V_{BG}$ conditions for Device #1 and Device #2 (see Supplementary Fig. 3 and Supplementary Fig. 4), we extract average $\Delta_{QPC}$ values of 24.5$\pm$1.3 meV and 29.8$\pm$2.5 meV, respectively (Supplementary Table 1 and 2).

| $V_{BG}$ (V) | 70 | 71 | 72 |
|---|---|---|---|
| $\Delta V_{SD}/2$ (V) | 29.83 | 28.88 | 27.57 |

**Supplementary Table 1**. $\Delta V_{SD}/2$ values of Device #1 for examined $V_{BG}$ conditions.

| $V_{BG}$ (V) | 70 | 71 | 72 | 73 | 74 | 75 | 76 | 77 | 78 | 79 | 80 |
|---|---|---|---|---|---|---|---|---|---|---|---|
| $\Delta V_{SD}/2$ (V) | 29.83 | 28.88 | 27.57 | 30.5 | 33.63 | 33.57 | 33.25 | 28.7 | 27.05 | 28.2 | 27.35 |

**Supplementary Table 2**. $\Delta V_{SD}/2$ values of Device #2 for examined $V_{BG}$ conditions.

**Intrinsic mobility**

It has been known that the two-terminal mobility ($\mu^{2T}$) severely underestimates the true intrinsic channel mobility due to the contact resistance at the metal-TMDC interfaces. To reasonably estimate the intrinsic mobility of our device, we refer to our group's previous charge transport study on multilayer $MoS_2$ devices with identical Indium (In) contacts[29]. In that study, the intrinsic mobility of a 6-layer $MoS_2$ obtained from four-probe measurements was nearly three times higher (~3200 $cm^2V^{-1}s^{-1}$) than the uncorrected two-probe mobility (~1200 $cm^2V^{-1}s^{-1}$) in the low-temperature regime ($T < 20$ K). We suggest that applying this factor of 3 to our current $WSe_2$ device represents a highly rigorous and conservative lower bound. Because indium metal is evaporated at a relatively low thermal energy (~ 530°C), it forms an ultraclean van der Waals contact that minimizes defect-induced Fermi-level pinning, thereby closely following the Schottky-Mott rule[29]. The electron affinity of multilayer $WSe_2$ (~3.6 eV) is significantly smaller than that of $MoS_2$ (~4.1 eV). Given the work function of Indium (~4.1 eV), the In Fermi level in $WSe_2$ resides deeper within the bandgap compared to the near-Ohmic In/$MoS_2$ interface. This band alignment inevitably results in a higher Schottky barrier and a correspondingly larger contact resistance for In/$WSe_2$, compared to those at In/$MoS_2$. This interpretation is also supported by Ref. [30]. Since the mobility factor between two-terminal configurations ($\mu^{2T}$) and intrinsic one ($\mu^{est}$) was ~3 for the highly transparent In/$MoS_2$ contacts, thus, we suggest that the factor for In/$WSe_2$ is not less than 3, at least. Based on this discussion, the mobility of Device #1 and Device #2 are estimated as $\mu = 1680$ and $600$ $cm^2V^{-1}s^{-1}$ with $\mu^{2T} = 560$ and $200$ $cm^2V^{-1}s^{-1}$, respectively.

**Mean free path ($l_e$)**

For Device #1, to quantitatively determine the transport regime, we evaluated the relevant length scales at $V_{BG}$ = 60 V and $T$ = 4 K, based on data in Fig. 3a. The applied effective gate voltage is $\Delta V_{BG}$ = $V_{BG} - V_{th}$ = 25 V with a threshold voltage, $V_{th}$ = 35 V (see Fig. 1c). Based on the dielectric thicknesses (500 nm thick $SiO_2$ and 17.7 nm thick h-BN), the 2D carrier density is estimated to be $n \approx 1.04\times10^{12}$ $cm^{-2}$. Due to the strong electric field and the SVL coupling in our device, the spin and valley degeneracies are fully lifted in the first few subbands.

The Fermi wave vector and Fermi wavelength are obtained by the relation $k_F = \sqrt{4\pi n}$ a Fermi wavelength of $\lambda_F = 2\pi/k_F$, respectively, where we assume that the spin and valley degeneracies are fully lifted in the first few subbands, due to the strong electric field and the SVL coupling in our device. Then, a mean free path is obtained by the relation $l_e = \hbar k_F \mu/e$. Therefore, we obtain $k_F \approx 3.61\times10^8$ $m^{-1}$, which yields a Fermi wavelength of $\lambda_F \approx 17.4$ nm. Finally, with $\mu \approx 1680$ $cm^2V^{-1}s^{-1}$, we get $l_e \approx 40$ nm.

For Device #2, to quantitatively determine the transport regime, we evaluated the relevant length scales at $V_{BG}$ = 80 V and $T$ = 3 K, based on data in Fig. 4g. The applied effective gate voltage is $\Delta V_{BG}$ = $V_{BG} - V_{th}$ = 34 V with a threshold voltage, $V_{th}$ = 46 V (see Fig. 1d). Based on the dielectric thicknesses (500 nm thick $SiO_2$ and 34 nm thick $h$-BN), we get $n \approx 1.37\times10^{12}$ $cm^{-2}$. Then, we obtain $k_F \approx 4.15\times10^8$ $m^{-1}$, $\lambda_F \approx 15.1$ nm, $l_e \approx 16.4$ nm with $\mu \approx 600$ $cm^2V^{-1}s^{-1}$. Supplementary Table 3 and 4 show experimental parameters of devices and extracted length scales. Supplementary Table 3 summarizes the experimental parameters of the two devices.

| | $V_{th}$ (V) | $\Delta V_{BG}$ (V) | $n$ ($cm^{-2}$) | $\mu^{2T}$ ($cm^2V^{-1}s^{-1}$) | $\mu$ ($cm^2V^{-1}s^{-1}$) | $\Delta_{qpc}$ (meV) |
|---|---|---|---|---|---|---|
| Device #1 | 35 | 25 | $1.04\times10^{12}$ | 560 | 1680 | 24.5 |

| Device #2 | 46 | 34 | $1.37\times10^{12}$ | 200 | 600 | 29.8 |
|---|---|---|---|---|---|---|

Supplementary Table 3. Experimentally obtained physical parameters of the two devices.

**Transverse confinement and effective width ($W_{\text{eff}}$)**

Using the Büttiker saddle-point potential model, the transverse confinement created by the electrostatic split gates is approximated as a harmonic oscillator, $V(y) = \frac{1}{2}m_e^*\omega_y^2y^2$, where $m_e^*$ is an effective mass of electron ($= 0.4m_e$) and $\omega_y$ represents the transverse confinement frequency. In this framework, the energy spacing between consecutive 1D spatial subbands is fundamentally governed by this potential, following the relation $\Delta_{\text{QPC}} = \hbar\omega_y$. The effective channel width ($W_{\text{eff}}$) can be analytically extracted as: $W_{\text{eff}} = 2\frac{v_F}{\omega_y}$, where $v_F = \frac{\hbar k_F}{m_e^*}$. Supplementary Table 4 shows $W_{\text{eff}}$ values of the two devices.

**Longitudinal barrier curvature ($\omega_x$) and effective channel length ($L_{\text{eff}}$)**

The longitudinal potential along the transport direction is modeled as an inverted parabola, $V(x) = -\frac{1}{2}m_e^*\omega_x^2x^2$, where $\omega_x$ represents the barrier curvature. In characteristic split-gate QPC geometries, the electrostatic potential varies much more gradually along the transport direction than in the transverse direction. This structural asymmetry is well-described by the empirical relation $\omega_x \approx (0.2\sim0.5)\omega_y$[31]. The effective channel length ($L_{\text{eff}}$) is fundamentally characterized by the spatial extent of this parabolic barrier evaluated at the Fermi energy. Based on the classical turning points of the inverted parabolic potential, the effective channel length ($L_{\text{eff}}$) is estimated as: $L_{\text{eff}} = W_{\text{eff}}\left(\frac{\omega_y}{\omega_x}\right)$. Finally, Table R1-2 summarizes $k_{\text{F}}$ and length scales of the two devices. Supplementary Table 4 shows $L_{\text{eff}}$ values of the two devices.

| | $k_{\text{F}}$ ($\text{m}^{-1}$) | $\lambda_{\text{F}}$ (nm) | $l_e$ (nm) | $W_{\text{eff}}$ (nm) | $L_{\text{eff}}$ (nm) |
|---|---|---|---|---|---|
| Device#1 | $3.61\times10^8$ | 17.4 | 40 | 5.6 | 11.2 ~ 28 |
| Device#2 | $4.15\times10^8$ | 15.1 | 16.4 | 5.3 | 10.6 ~ 26.5 |

Supplementary Table 4. Extracted $k_{\text{F}}$ and length scales of the two devices.

**Phase coherence length**

While widespread magneto-transport studies reporting long phase coherence lengths ($l_{\phi}$ ~ 100 - 300 nm) in $WSe_2$ predominantly focus on p-type transport[10], the fundamental physics applies equally to our n-type multilayer device. In multilayer $WSe_2$, the conduction band minimum is at the Q valley. The effective mass of these Q-valley electrons ($m_e^* \approx 0.4m_e$) is highly comparable to that of the K-valley holes. According to the Altshuler-Aronov-Khmelnitsky theory for 2D systems[32], this similarity in effective mass yields comparable diffusion coefficients and Nyquist electron-electron dephasing rates. Consequently, the electron phase coherence length remains on the order of hundreds of nanometers, strictly satisfying the condition $l_{\phi} \gg L_{\mathrm{eff}}$. Because coherence is perfectly preserved across the short channel, the local disorder induces strong coherent resonant backscattering. This quantum interference effectively reduces the forward transmission strictly below unity ($Tr < 1$), causing the ideal flat plateaus to collapse into the pronounced peak-dip conductance fluctuations observed in Fig. 2d, 4a, while the transverse 1D subband spacing remains robust.

As the temperature increases to 20 K in Fig. 4a, thermal dephasing sharply reduces $l_{\phi}$, suppressing the quantum interference effects and smoothing out the resonant peaks. Because our split-gate design induces a large subband gap ($\Delta_{\mathrm{QPC}} \approx$ 30 meV) that overcomes the thermal energy ($k_B T \approx$ 1.7 meV at 20 K), the underlying 1D QPC quantization remains perfectly protected against thermal smearing. This is why the pristine step-like quantized plateaus apparently emerge at elevated temperatures with $Tr < 1$.

**Transport regime of Device #1**

To comprehensively define the electron transport regime of Device #1, we evaluate the hierarchy of the fundamental length scales: the Fermi wavelength ($\lambda_F$), the mean free path ($l_e$), and the effective channel dimensions ($W_{\mathrm{eff}}$ and $L_{\mathrm{eff}}$). First, the estimated effective channel width ($W_{\mathrm{eff}} \approx$ 5.6 nm) is comparable to half of the Fermi wavelength ($\lambda_F/2 \approx$ 8.7 nm), which allows the fundamental 1D QPC confinement subband to be cleanly isolated and observed, directly leading to the distinct subband energy gap.

Second, the effective channel length evaluated from the anisotropic saddle-point model ($L_{\mathrm{eff}} \approx$ 11.2 - 28 nm) is distinctly shorter than the momentum-relaxing mean free path of the 2DEG ($l_e \approx$ 40 nm). While the condition $L_{\text{eff}} \leq l_e$ generally implies highly ballistic transport, the specific transport behavior observed at low temperatures ($T$ < 10 K) requires a more rigorous classification (see Fig. R1-1). At temperatures below 10 K, we observe pronounced peak-dip features superimposed on the quantized conductance plateaus. In this low-temperature regime, the phase coherence length ($l_\phi$) extends significantly, far exceeding the channel length ($l_\phi \gg L_{\text{eff}}$). Consequently, even minimal residual impurities or local potential fluctuations near the channel can act as phase-coherent scattering centers. The observed peak-dip phenomena are clear manifestations of quantum resonant backscattering, where electrons undergo multiple coherent reflections, leading to quantum interference that slightly degrades the perfect forward transmission ($Tr$ < 1). Therefore, incorporating both the length-scale hierarchy and the low-temperature interference effects, we classify the transport mechanism of Device #1 as operating in the phase-coherent quasi-ballistic regime.

**Transport regime of Device #2**

For Device #2, we obtain the fundamental length scales governing the QPC channel: the Fermi wavelength ($\lambda_F \approx$ 15.1 nm), the mean free path ($l_e \approx$16.4 nm), and the effective channel dimensions ($W_{\text{eff}} \approx$ 5.3 nm and $L_{\text{eff}} \approx$ 10.6 – 26.5 nm). First, the estimated effective $W_{\text{eff}}$ is comparable to half of the Fermi wavelength ($\lambda_F/2 \approx$ 7.6 nm), as Device#1. On the other hand, $L_{\text{eff}}$ is comparable to $l_e$. Under this specific condition ($L_{\text{eff}} \approx l_e$), electrons traversing the QPC constriction have a non-negligible probability of undergoing one or a few impurity scattering events before exiting the channel. Because these scattering events inevitably reduce the forward transmission probability below unity ($Tr$ < 1), the transport mechanism of Device #2 is unambiguously defined as operating in the quasi-ballistic regime.

## Supplementary Note 4. $\delta V_{SD}^{-}$-$V_{BG}$ curve for Device #1 and $\delta V_{SD}^{-}$-$D$ curves for Device #2

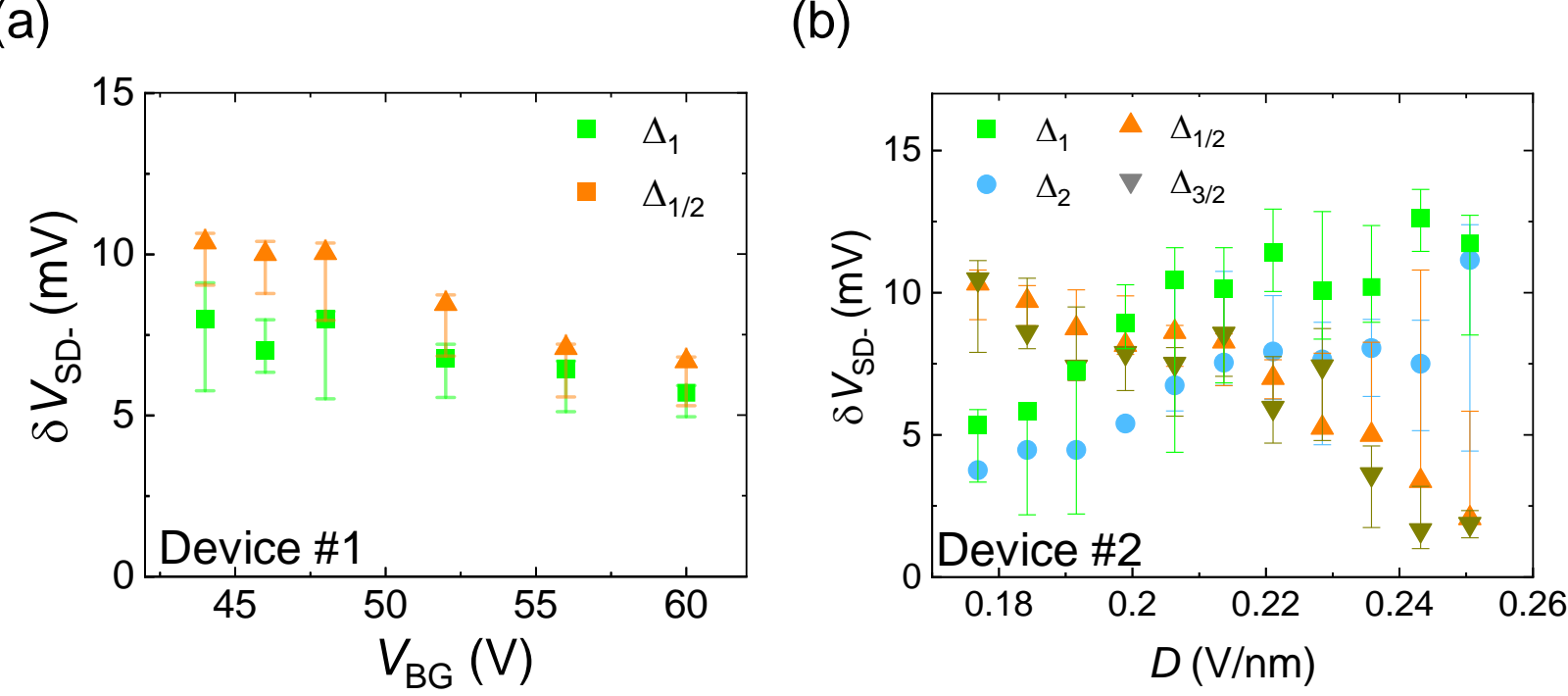

**Supplementary Figure 6.** (a) $\delta V_{SD}^{-}$-$V_{BG}$ curve for Device #1. (b) $\delta V_{SD}^{-}$-$D$ curves for Device #2.

## Supplementary Note 5. The effect of layer thickness with respect to observation of the SVL-induced spin splitting

### Effectiveness of the out-of-plane electric field (thickness and screening effect)

The out-of-plane electric field ($E_{\perp}$) generated by the back-gate does not effectively penetrate all 14 layers. In multilayer transition metal dichalcogenides (TMDCs), the gate-induced electric field is subject to strong Thomas-Fermi screening along the out-of-plane ($c$-axis) direction. Both theoretical calculations and experimental studies have consistently shown that the electrostatic screening length in TMDCs is typically around 1 to 3 atomic layers (~0.7 to 2 nm) [33,34]. Consequently, the gate-induced charge carriers are heavily localized in the bottom-most 1 to 3 layers closest to the gate dielectric, leaving the upper layers largely un-gated or only weakly coupled. In the 14-layer device (~10 nm thick), this strong screening creates a highly inhomogeneous charge distribution. The upper layers act as an un-gated or weakly gated parasitic parallel channel. Furthermore, the thick 14-layer crystal results in a much smaller energy spacing between the out-of-plane 2D subbands, induced by the quantum confinement effect along the z-axis[35]. To provide a highly quantitative perspective based on a standard 1D quantum well model for a layer thickness ($t$), we evaluated the out-of-plane subband spacing ($\Delta E_{1,2} = 3\pi^2\hbar^2/2m_{e,z}^{*}t^2$) using the material parameters of multilayer $WSe_2$. In multilayer $WSe_2$, the conduction band minimum is located at the Q valley, where the out-of-plane effective mass for electrons is approximately $m_{e,z}^{*} \approx 0.64m_0$[36]. Applying this effective mass of $0.64m_0$, the energy spacing between the subbands is drastically different for the two devices. For the 14-layer device with a physical thickness of 9.8 nm, we get $\Delta E$ ~ 18 meV. This energy spacing is easily overcome by local disorder potential fluctuations, fringing-field variations, and lifetime broadening within the QPC constriction. Therefore, the 14-layer device inherently suffers from severe subband mixing, which fundamentally smears out the discrete 1D quantization. In contrast, for the 5-layer device with a physical thickness of 3.5 nm, the subband spacing is calculated to be robustly large, $\Delta E \approx$ 140 meV. This massive energy barrier completely suppresses inter-subband scattering and rigorously guarantees pure single-subband ballistic transport. Furthermore, the 5-layer device is much closer to the electrostatic screening limit. A significantly larger fraction of the channel is effectively controlled by the gate, and the subband spacing is wider, which strongly suppresses inter-layer parasitic transport and allows the robust observation of 1D conductance quantization.

### Proof of interlayer coupling in multi-layer $WSe_2$

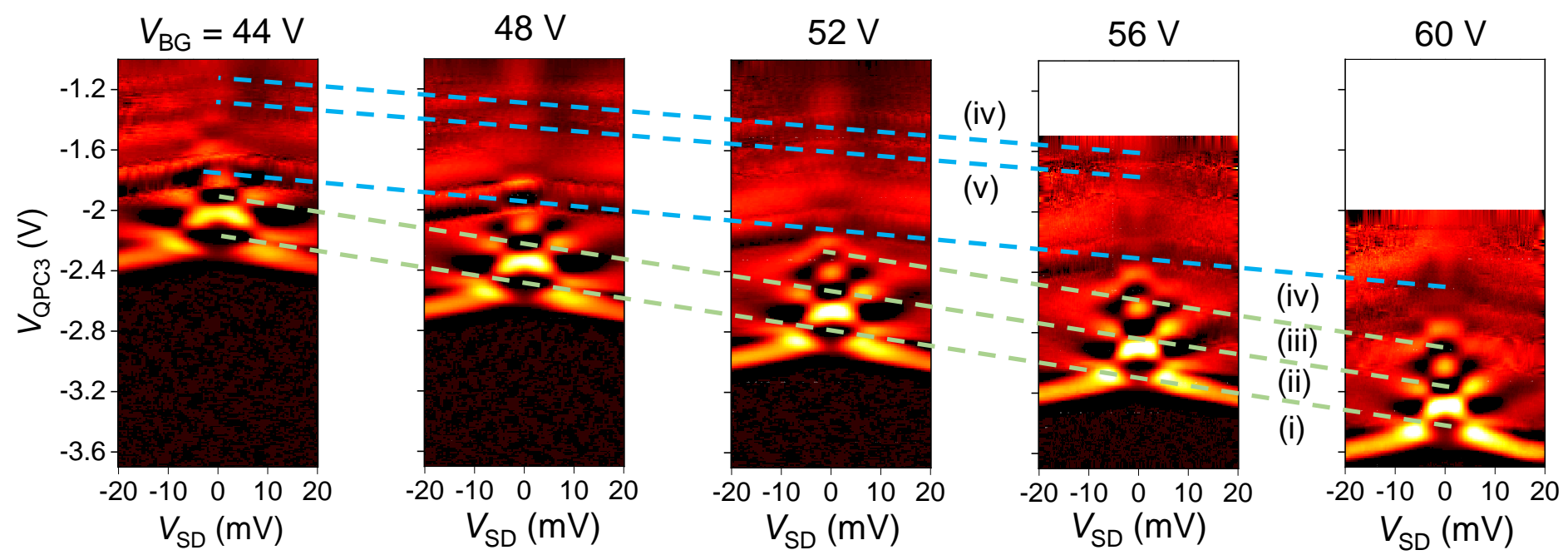

**Supplementary Figure 7.** Extended $\mathrm{d}G/\mathrm{d}V_{QPC3}$ map, based on data Fig. 3a of Device #1.

Supplementary Figure 7 shows the extended finite-bias transconductance maps of plots in Fig. 3a of the revised version, except $V_{BG}$ = 46 V for Device #1. While the fundamental subband diamonds are pristine (indicated by dashed green lines), the higher-energy regions (dashed blue lines) become highly complex, compared to simple 1D subbands.

The degradation of pristine 1D quantization at higher conductance is a physically expected consequence of our specific 14-layer $WSe_2$ device architecture. As already discussed in the main text (regarding the anomalous $V_{SD}$ dependence of $\Delta_1$ in Fig. 3), the multi-layer nature of our device inherently introduces interlayer coupling. While the fundamental subbands ($\Delta_{1/2}$ and $\Delta_1$) are strongly confined to the few bottom layers via spin-valley-layer (SVL) coupling, yielding clear quantized plateaus, the strong finite $V_{SD}$ conditions begin to induce subtle upper-layer interference even at the first subband level.

We present the $V_{BG}$-dependent transconductance maps (Supplementary Fig. 7) and a quantitative analysis of the capacitive coupling ratios (Supplementary Fig. 8). What Supplementary Fig. 7 and 8 demonstrate is how this interlayer coupling transits from a subtle perturbation at the fundamental level to a dominant force at higher subbands. We tracked the gate-voltage positions of the subband features and the anomalous background structures as a function of $V_{BG}$ (Supplementary Fig. 8), revealing two quantitatively distinct regimes based on their slopes ($\Delta V_{QPC3}/\Delta V_{BG}$):

**1) Intrinsic 1D subbands (green symbols, i-iii):** The primary subbands exhibit a steep and highly consistent slope of $\Delta V_{\mathrm{QPC3}}/\Delta V_{\mathrm{BG}} \sim -0.075$. This relatively large magnitude signifies robust electrostatic SVL coupling to the back-gate, confirming that the core 1D transport is governed by states at the few bottom $WSe_2$ layers.

**2) Parasitic upper-layer states (blue symbols, iv-vi):** In contrast, the anomalous background features exhibit a significantly shallower slope (in a range of $\Delta V_{\mathrm{QPC3}}/\Delta V_{\mathrm{BG}} = -0.037 \sim -0.046$). This nearly two-fold reduction in capacitive coupling is quantitative evidence for electrostatic screening. These shallow-sloped states reside spatially further away from the back gate, specifically, in the upper layers of the 14-layer flake.

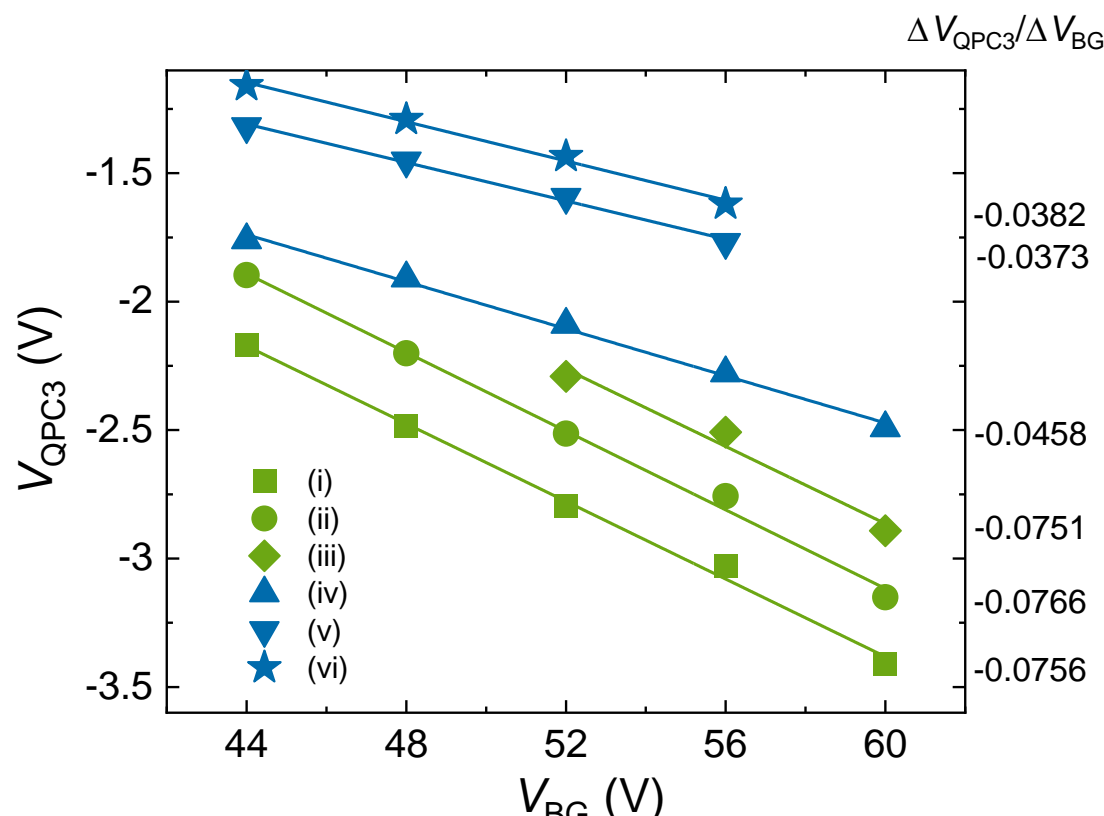

**Supplementary Figure 8.** Capacitive coupling ratios of dashed green and blue lines in Supplementary Figure 7.

As the QPC channel is opened to reach higher conductance levels (> $2e^2/h$), the confinement potential widens, and the electronic wavefunctions could expand vertically. Consequently, the transport strongly hybridizes with these weakly coupled, parasitic upper-layer states. Because this severe multi-layer interference completely washes out the pristine 1D nature at higher energies, it physically prevents the formation of higher-order plateaus. Therefore, we confidently restrict our primary quantitative analysis to the fundamental subband regimes, where SVL coupling successfully preserves the 1D quantum limit despite the multi-layer effects for Device #1.

**Role of even and odd layer numbers**

Regarding the parity effect, in pristine $WSe_2$, odd-numbered layers globally lack inversion symmetry, whereas even-numbered layers possess global inversion symmetry. If the device were in a pristine, un-gated state, this parity would fundamentally dictate the presence (in odd layers) or global absence (in even layers) of macroscopic spin-valley locking. Indeed, as demonstrated by magneto-transport studies in few-layer TMDCs[37], this parity effect directly governs whether the system exhibits a valley Zeeman effect (odd layers) or a spin Zeeman effect (even layers). However, in our heavily gated QPC transport regime, the even/odd parity does not play a dominant role. The application of a strong out-of-plane $V_{BG}$ introduces a strong structural inversion asymmetry. Because the electric field and the charge carriers are heavily concentrated at the bottom-most layers (due to the screening effect discussed above), the local inversion symmetry at these bottom layers is broken by the perpendicular electric field, regardless of the global even or odd parity of the entire flake. It has been well established that such out-of-plane electric fields effectively restore and tune the spin-valley locking even in inversion-symmetric even-layer TMDCs[38]. Therefore, the strong spin-valley-layer (SVL) coupling required for our QPC operation is induced primarily by the strong external electric field, making the pristine global even/odd layer count practically irrelevant to the observed QPC transport behavior.

**Supplementary Note 6. 2D map of G as a function of $V_{QPC2}$ and $V_{BG}$**

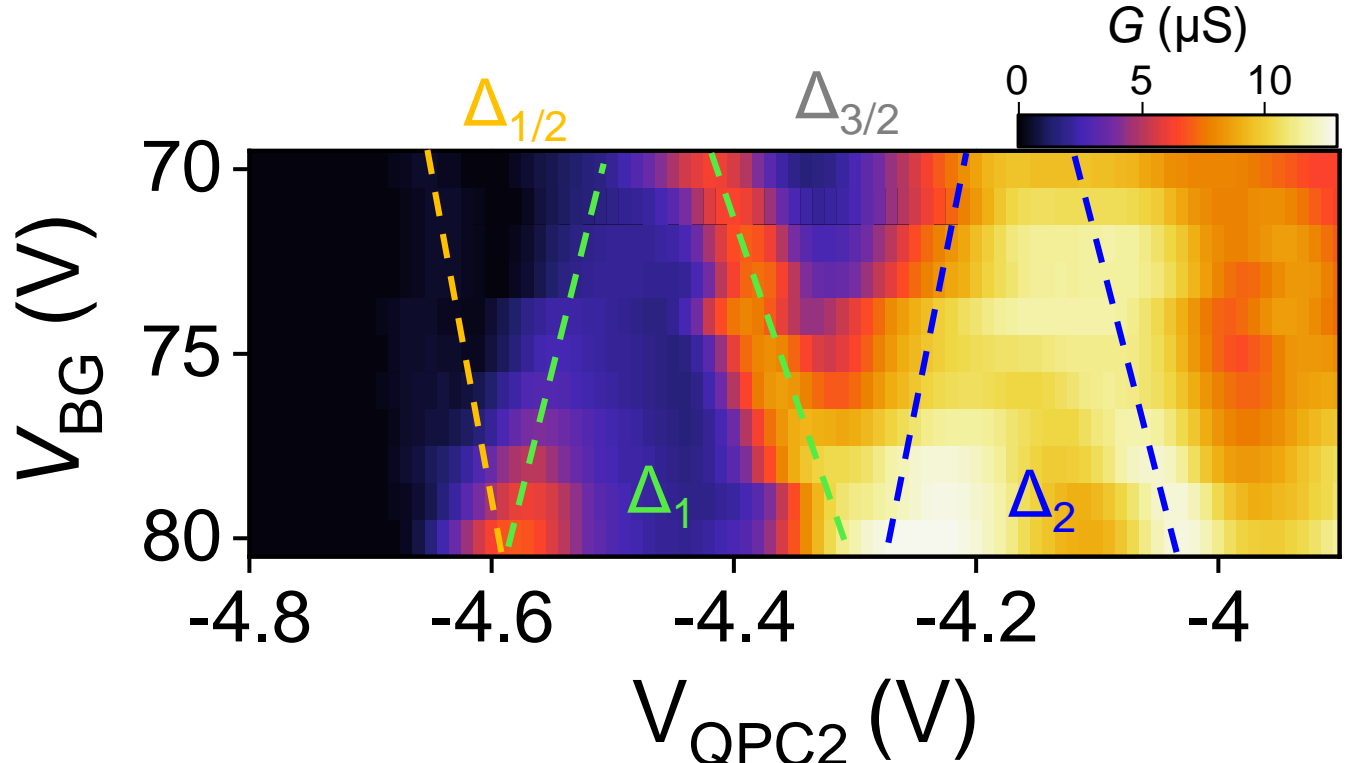

**Supplementary Figure 9.** Conductance ($G$) as a function of $V_{\mathrm{QPC2}}$ and $V_{\mathrm{BG}}$ of Device #2.

**Supplementary Note 7. d$I$/d$V_{\mathrm{QPC2}}$ maps at $V_{\mathrm{BG}}$ = 70, 80, 90 and 130 V for Device #2**

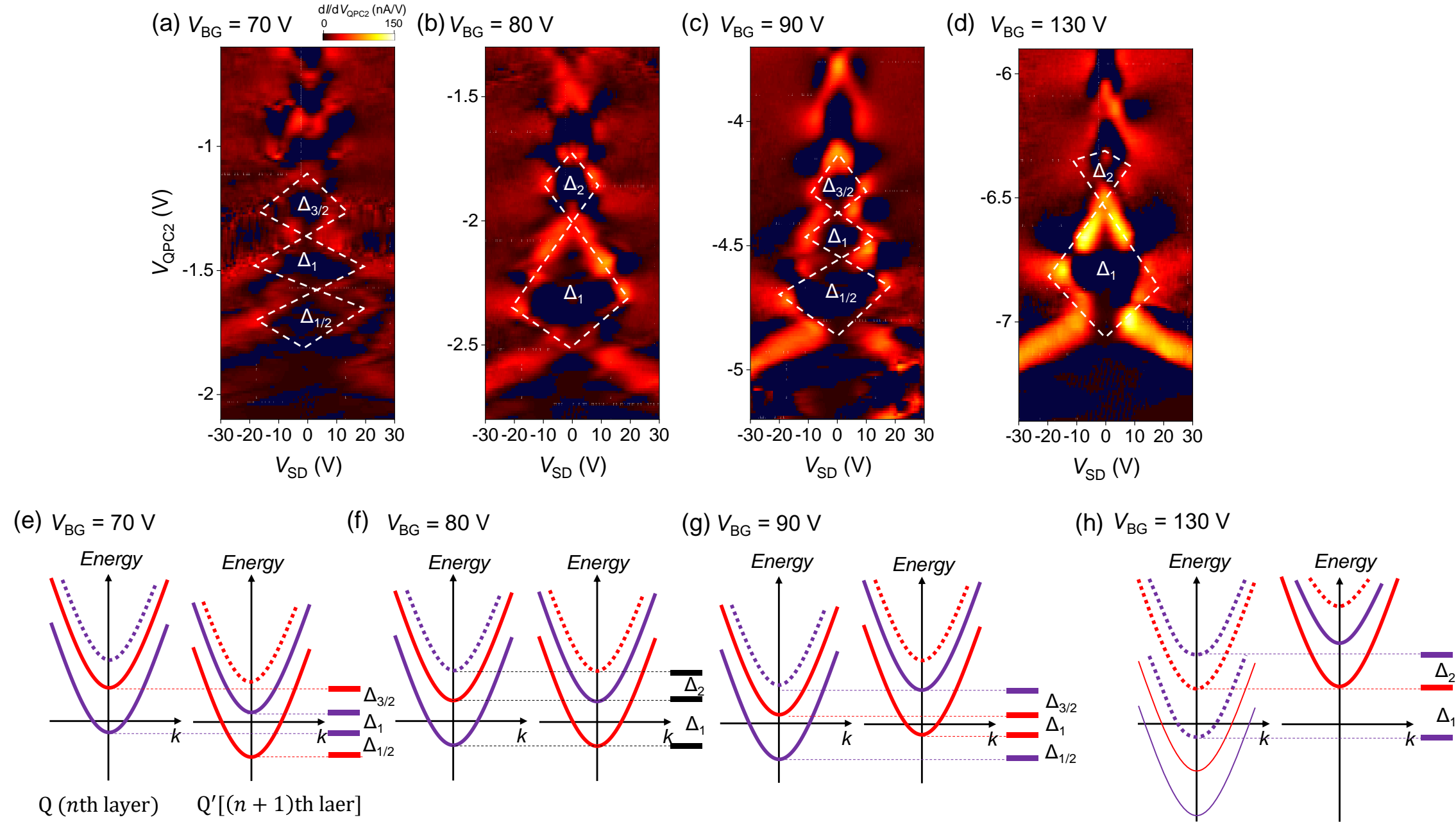

**Supplementary Figure 10.** (a)-(d) d$I$/d$V_{\mathrm{QPC2}}$ maps at $V_{\mathrm{BG}}$ = 70, 80, 90 and 130 V of device#2. (e)-(h) SVL-coupling induced band shifts in Q and Q' valleys in the conduction band, located in $n$th and ($n$+1) layer, respectively, at corresponding $V_{\mathrm{BG}}$ values.

**Supplementary Note 8. d$G$/d$V_{\mathrm{QPC2}}$ – $V_{\mathrm{QPC2}}$ curve at $V_{\mathrm{BG}}$ = 80 V of Device #2**

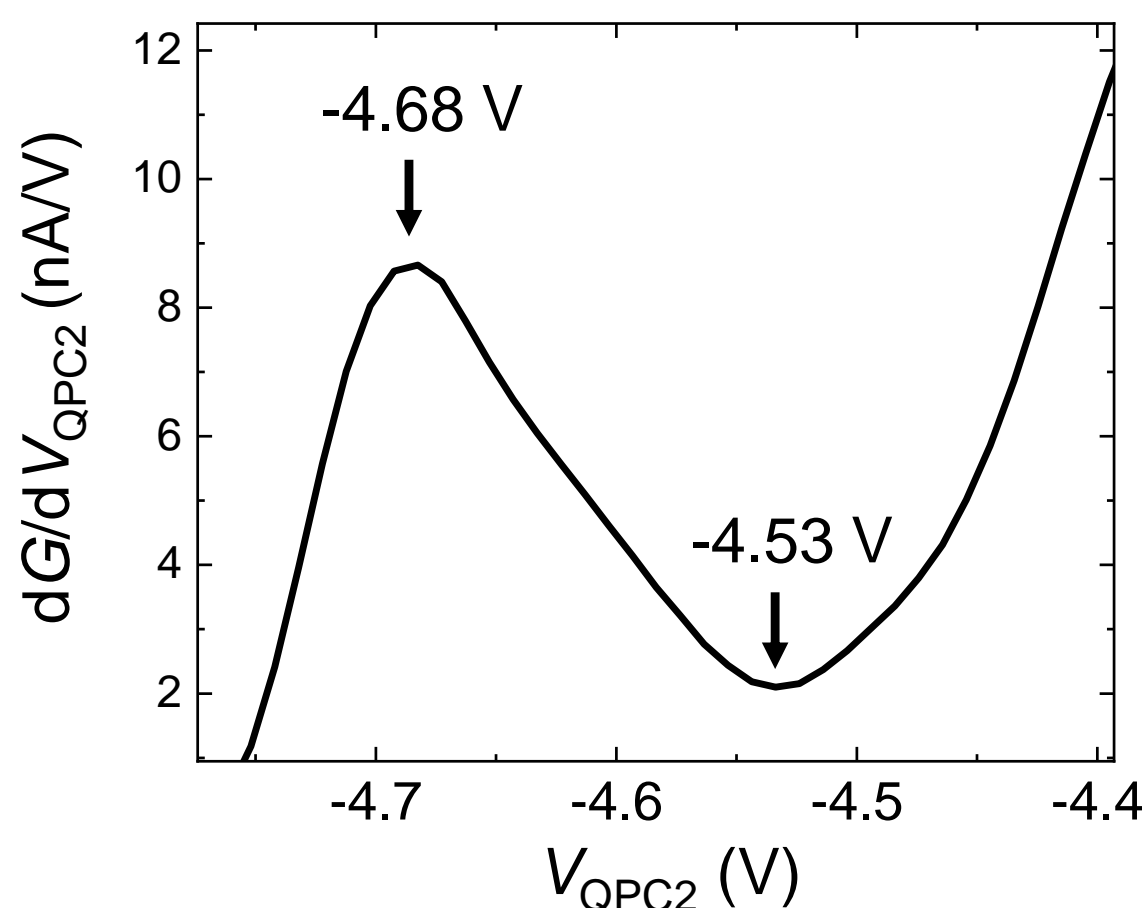

**Supplementary Figure 11.** d$G$/d$V_{\mathrm{QPC2}}$- $V_{\mathrm{QPC2}}$ curve at $V_{\mathrm{BG}}$ = 80 V of Device #2

## Supplementary Note 9. Noise-bounded error bar evaluation procedure

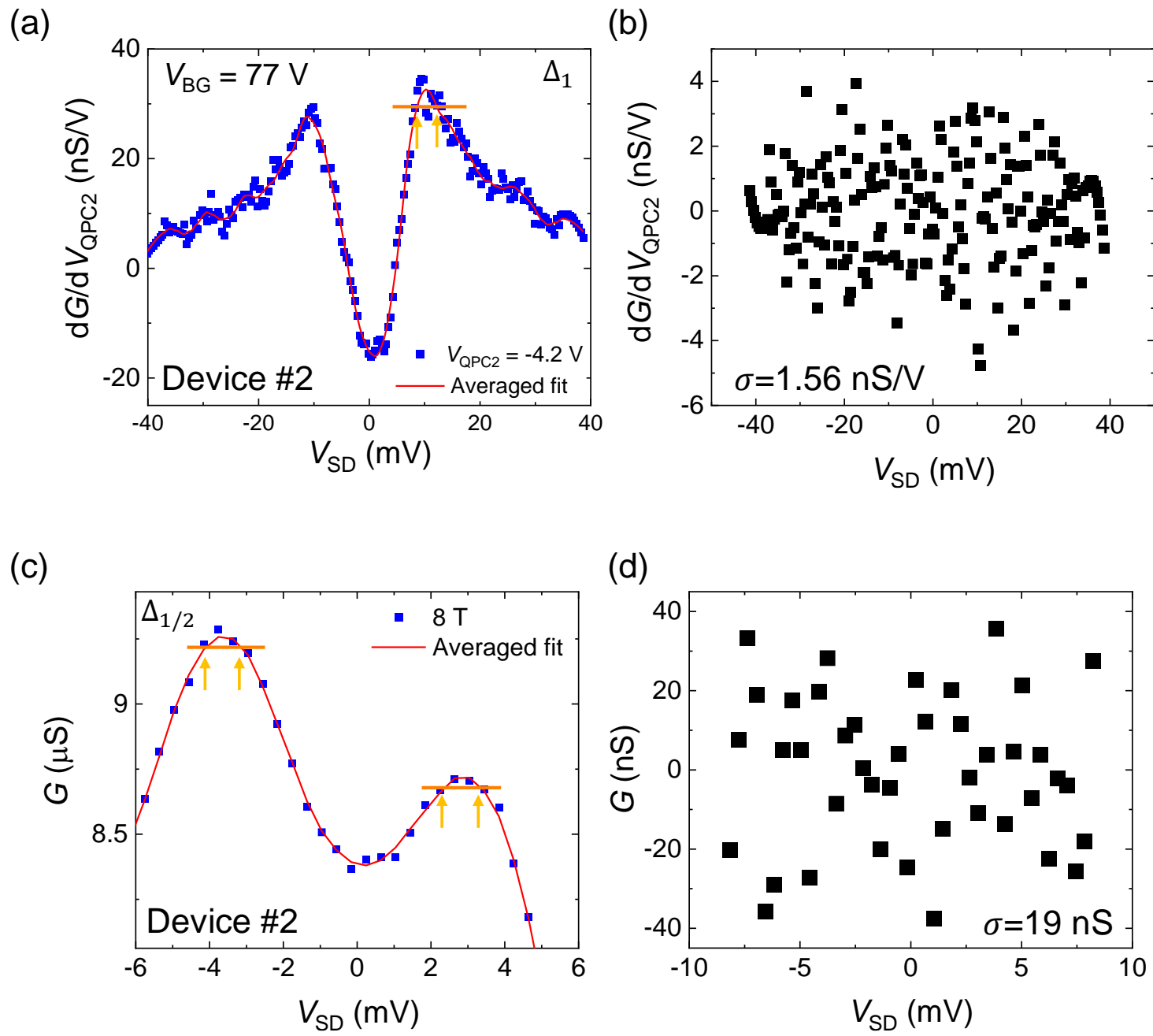

**Supplementary Figure 12.** Noise-bounded error bar evaluation procedure.

Supplementary Fig. 12 shows the noise-bounded error bar evaluation procedure. The scatter plot in Supplementary Fig. 12a displays d$G$/d$V_{\mathrm{QPC2}}$ as a function of $V_{\mathrm{SD}}$, measured at $V_{\mathrm{QPC2}}$ = −4.2 V and $V_{\mathrm{BG}}$ = 77 V for Device #2 (corresponding to the data in Supplementary Fig. 4). To evaluate the noise level of these measurements, we applied a smoothed fitting to the raw data, represented by the red curve. Supplementary Fig. 12b plots the residuals (the difference between the raw data and the fitting curve), yielding a standard deviation of $\sigma$ = 1.56 nS/V. A $2\sigma$ threshold was then utilized to determine the noise-bounded error bars. For instance, we defined horizontal orange bars across both d$G$/d$V_{\mathrm{QPC2}}$ peaks in Supplementary Fig. 12a, positioned 3.1 nS/V ($2\sigma$) below the maximum peak amplitudes. The intersections of these bars with the data profile establish the lower and upper bounds of the error bars for $\delta V_{SD}^{+}$, as indicated by the two orange arrows. The error bars presented in Fig. 3b,c and Fig. 4h, i of the main text were derived following this procedure. Similarly, the error bars in Fig. 6d were evaluated based on the procedure detailed in Supplementary Fig. 12c, d.

## Reference for Supplementary Information